%% file: main.tex
\documentclass[runningheads]{llncs}

\input{preamble}

\input{meta}

\begin{document}

\maketitle

\begin{abstract}
\input{sections/00-abstract-1.tex}

\end{abstract}

\iffalse
Conference and track fit.* Please, provide a justification of why your submission fits the scope of the conference and of the research track.

We consider the interplay of OWL and SHACL, both recommendations of the W3C. We explore how these formalisms can be combined by using SHACL only, an interesting task that has barely been addressed before. About the theory of the combination, less than a handful, but well-received papers have been written. The practical side of the problem is completely novel, as far as we are aware. We address both the theory, for a new fragment, as the practical side of the problem.
\fi

\input{sections/01-introduction}

\input{sections/02-preliminaries}

\input{sections/03-warm-up}
\input{sections/04-rewriting}
\input{sections/08-implementation}

\input{sections/08b-towards-full-OWL-EL}
\input{sections/09-conclusions-and-outlook}

\input{sections/acknowledgements}
\clearpage
\input{sections/supplemental-material-statement}
\input{sections/use-of-gen-ai-statement}
\bibliographystyle{splncs04}
\bibliography{bibliography}

\clearpage

\input{sections/appendix}

\end{document}

%% file: preamble.tex
\usepackage{graphicx} % Required for inserting images
\usepackage{amsmath}
\usepackage{amssymb}
\usepackage{mathtools}
\usepackage{stmaryrd}
\usepackage[dvipsnames]{xcolor}
\usepackage{soul}
\usepackage{url}
\usepackage[hidelinks]{hyperref}
\usepackage[small]{caption}
\usepackage{graphicx}
\usepackage{booktabs}
\usepackage{tabularx}
\usepackage[ruled,vlined,noend]{algorithm2e}
\usepackage{algpseudocode}
\usepackage{subfig}
\usepackage{todonotes}
\usepackage{xspace}
\usepackage{paralist}
\usepackage[mathscr]{euscript}
\usepackage{multicol}
\usepackage{listings}
\lstdefinestyle{turtle}{%
  basicstyle=\ttfamily\tiny,
  columns=flexible,
  keepspaces=true,
  breaklines=true,
  breakatwhitespace=false,
  numbers=left,
  numberstyle=\tiny\color{Gray},
  numbersep=4pt,
  frame=single,
  framesep=2pt,
  xleftmargin=14pt,
  xrightmargin=4pt,
  alsoletter={@:-},
  morestring=[bd]{<}{>},
  stringstyle=\color{black},
  morecomment=[l]{\#},
  commentstyle=\color{Gray}\itshape,
  escapeinside={(*@}{@*)},
  keywords=[1]{@prefix,a},
  keywordstyle=[1]\color{Plum}\bfseries,
  keywords=[2]{sh:NodeShape,sh:property,sh:path,sh:class,sh:minCount,
    sh:node,sh:or,sh:not,sh:inversePath,sh:alternativePath,
    sh:zeroOrMorePath,sh:qualifiedMinCount,sh:qualifiedValueShape,
    sh:targetClass,sh:targetObjectsOf,sh:targetSubjectsOf},
  keywordstyle=[2]\color{NavyBlue}\bfseries,
  keywords=[3]{rdfs:subClassOf,rdfs:subPropertyOf,rdfs:domain,rdfs:range,
    owl:Class,owl:ObjectProperty,owl:onProperty,owl:someValuesFrom,
    owl:propertyChainAxiom},
  keywordstyle=[3]\color{OliveGreen}\bfseries,
}
\usepackage{tikz}
\usetikzlibrary{positioning, shapes.multipart, arrows.meta, backgrounds, fit, quotes}
\usepackage{xspace}
\usepackage{orcidlink}
\renewcommand{\orcidID}[1]{\,\orcidlink{#1}}

\usepackage{chngcntr}
\usepackage{bussproofs}

\hypersetup{
    colorlinks=true,
    linkcolor=black,
    urlcolor=blue,
    citecolor=black
}
\usepackage{cleveref}
\usepackage{multirow}
\newtheorem{df}{Definition}
\newtheorem{cor}{Corollary}
\newtheorem{lm}{Lemma}
\newtheorem{pr}{Proposition}
\spnewtheorem*{prooftheorem}{Proof of Theorem 1}{\itshape}{\rmfamily}
\newcommand*\imge{\raisebox{-0.17\baselineskip}{\includegraphics[height=0.81\baselineskip]{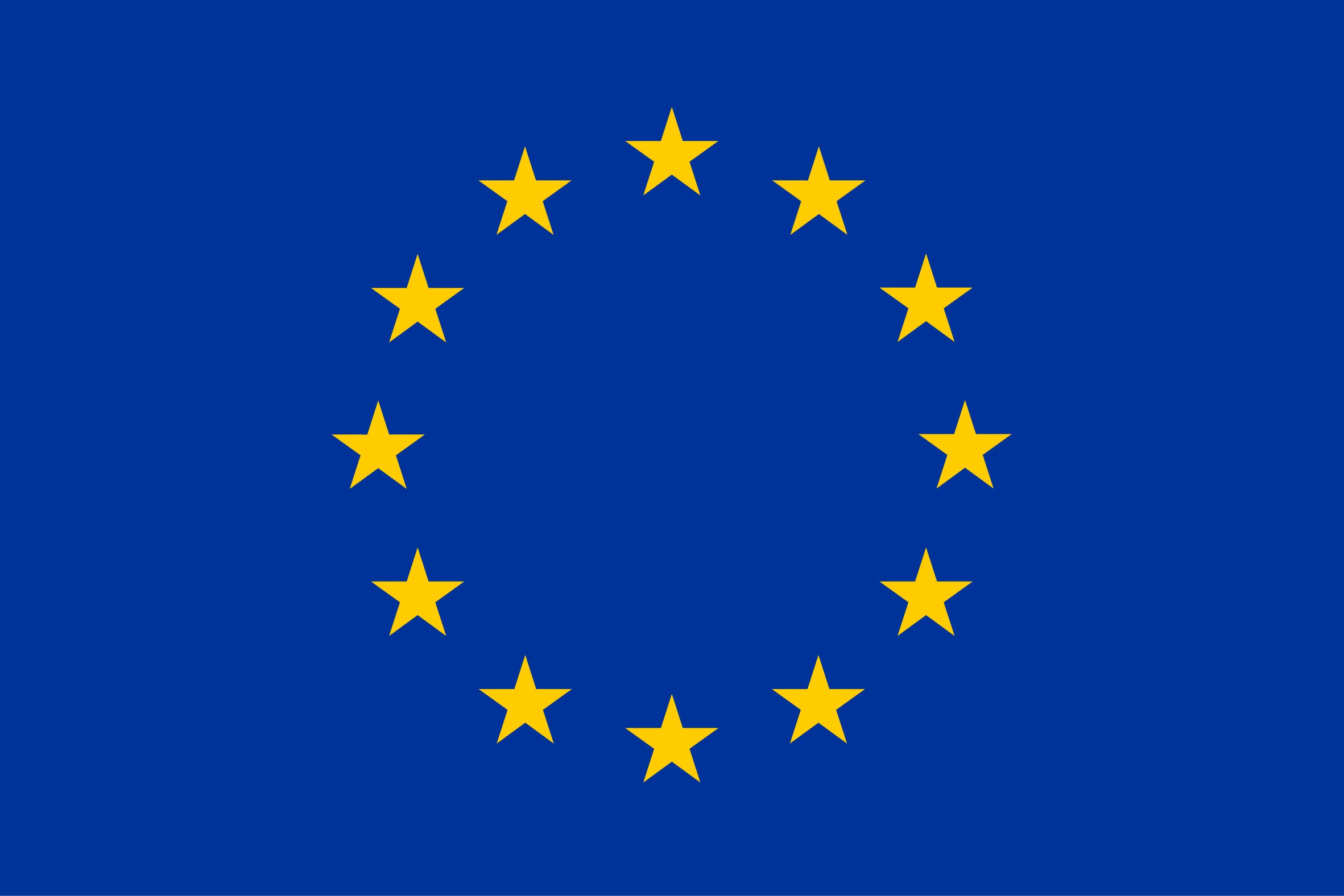}}}

\input{macros/dl}

\input{macros/pg}
\input{macros/shacl-rdf}

\input{macros/general}

\input{macros/tools}

%% file: macros/dl.tex
\newcommand{\sub}{\mathit{sub}}

\newcommand{\dom}[1]{\mathit{dom}(#1)}
\newcommand{\rang}[1]{\mathit{ran}(#1)}

\newcommand{\TBox}{\ensuremath{\mathcal{T}}\xspace}

\newcommand{\ABox}{\ensuremath{\mathcal{A}}\xspace}

%% file: macros/pg.tex
\newcommand{\set}[1]{\left\{#1\right\}}

\newcommand{\pg}[1]{\textcolor{teal}{PG: #1}}

\newcommand{\define}[1]{\textit{#1}}

\newcommand{\tuple}[1]{\left(#1\right)}
\newcommand{\guard}{\ \middle\vert\ }

\newcommand{\entails}{\models}

\newcommand{\val}[2]{\mathit{val}(#1, #2)}
\newcommand{\real}[2]{\mathit{cl}_{#2}(#1)}
\newcommand{\eqdef}{\mathrel{:=}}
\newcommand{\eqbnf}{\mathrel{::=}}
\newcommand{\rew}[2]{\mathit{rew}(#1,#2)}

\newcommand{\mathfun}[1]{\ensuremath{\mathit{#1}}}

\newcommand{\allShapes}{\mathfun{allShapes}}

\newcommand{\chains}{\mathfun{chains}}

\newcommand{\Base}{\mathfun{Base}}
\newcommand{\Dep}{\mathfun{Dep}}
\newcommand{\PathSigma}{\mathfun{\Sigma}}
\newcommand{\LPos}{L^{+}}
\newcommand{\LNeg}{L^{-}}

\newcommand{\param}[1]{\textcolor{teal}{\textit{#1}}}

%% file: macros/shacl-rdf.tex
\newcommand{\shape}{\ensuremath{\mathit{shape}}}

%% file: macros/general.tex
\newcommand{\eqhaak}{\sqsubseteq}

\newcommand{\A}{\mathcal{A}}
\newcommand{\T}{\mathcal{T}}

\newcommand{\G}{\mathcal{G}}

\newcommand{\C}{\mathcal{C}}

\newcommand{\I}{\mathcal{I}}
\newcommand{\J}{\mathcal{J}}

\newcommand{\elhi}{$\mathcal{ELHI}$}

\newcommand{\NC}{N_{C}}

\newcommand{\NR}{N_{R}}
\newcommand{\NA}{N_{A}}
\newcommand{\roles}{\overline{\NR}}
\newcommand{\NI}{N_{I}} 
\newcommand{\NS}{N_{S}}
\newcommand{\isa}{\sqsubseteq}

\newcommand{\elpp}{$\mathcal{EL}^{++}$\xspace}
\newcommand{\ran}{\mathit{ran}}

\newcommand{\exptime}{\textsc{ExpTime}\xspace}

\newcommand{\owlelm}{OWL EL$^-$\xspace}

\newcommand{\ggan}{\xspace\geq_n\!}
\newcommand{\ggae}{\xspace\geq_1\!}
\newcommand{\ggat}{\xspace\geq_2\!}

%% file: macros/tools.tex
\xspaceaddexceptions{+}

\newcommand{\owlawareshacl}{\textsc{ShacOWL}\xspace}
\newcommand{\hermit}{\textsc{HermiT}\xspace}
\newcommand{\jfact}{\textsc{JFact}\xspace}
\newcommand{\pellet}{\textsc{Pellet}\xspace}
\newcommand{\elk}{\textsc{ELK}\xspace}
\newcommand{\structuredreasoner}{\textsc{StructuralReasoner}\xspace}

\newcommand{\jena}{\textsc{Jena}\xspace}
\newcommand{\pyshacl}{\textsc{pySHACL}\xspace}
\newcommand{\isaitb}{\textsc{Isaitb}\xspace}
\newcommand{\topbraid}{\textsc{TopBraid}\xspace}

%% file: meta.tex
\newcommand{\ourtitle}{
Rewrite Once, Validate Anywhere: \\Producing OWL-Aware SHACL Constraints\\
(Extended Version)
}

\title{\ourtitle}

\newcommand{\shorttitle}{Producing OWL-Aware SHACL Constraints}
\titlerunning{\shorttitle}

\author{
	Anouk Oudshoorn%
	\inst{1}%
	\orcidID{0009-0006-4638-5948}%
	\and%
	Piotr Gorczyca%
	\inst{2}%
	\orcidID{0000-0002-6613-6061}%
	\and%
	Dörthe Arndt%
	\inst{2,3}%
	\orcidID{0000-0002-7401-8487}%
}

\authorrunning{A.Oudshoorn \and P. Gorczyca \and D. Arndt}

\institute{
	Institute of Logic and Computation, TU Wien, Austria%\\
	\and
	Computational Logic Group, TU Dresden,  Germany
	\and
	ScaDS.AI,
	Dresden/Leipzig, Germany \\  
	\email{anouk.oudshoorn@tuwien.ac.at}, \email{\{piotr.gorczyca,doerthe.arndt\}@tu-dresden.de}
}

%% file: sections/00-abstract-1.tex
% !TEX root = ../main.tex

The Shapes Constraint Language (SHACL) is a W3C recommendation to express syntactic constraints, called shapes, on RDF graphs.
SHACL validators  are used to test whether a given graph adheres to such a shape.
However, RDF graphs often come with OWL ontologies, whose implicit knowledge needs to be taken into account. This is classically handled by first applying 
reasoning and then performing the constraint checking on the results, often using different technologies which makes the process inefficient and vulnerable for mistakes. 

To overcome this, we 
propose to internalise the OWL axioms in the SHACL constraints; we construct a rewriting which takes as input both shapes and an OWL EL$^-$ ontology -- a fragment of OWL EL restricting the usage of existential restrictions -- and produces SHACL constraints. This output can then be evaluated by any validator supporting SHACL core regardless of its reasoning support, while yielding the same results as the traditional approach. The implementation of our translation is evaluated both against applying state-of-the-art reasoners and validators consecutively, as against validators with built-in reasoning support. 
For our benchmark, we show that our approach is in general more efficient in finding violations compared to the sequential approach, thus providing a powerful tool which simplifies combining reasoning with validation.

%% file: sections/01-introduction.tex
% !TEX root = ../main.tex
\section{Introduction}

The W3C recommendation SHACL \cite{shacl} makes it possible to specify syntactic constraints on RDF graphs, which may be checked for compliance. 
However, RDF graphs are not only syntactical constructs but also semantic statements. That is, 
they use OWL DL ontologies \cite{OWL} to define classes and predicates; they come with implicit knowledge. 
According to the specification, SHACL implementations only need to support subclass reasoning, all other reasoning support is optional
\cite[Section 1.5]{shacl}. 
As a result, the SHACL validation landscape is diverse. 
Many systems
only take the required subclass reasoning into account when performing constraint validation (e.g. \jena~\cite{jena}, \topbraid~\cite{topbraid-shacl}, and \isaitb~\cite{isaitb-shacl}), 
while others support OWL DL profiles (e.g., \pyshacl~\cite{pyshacl}). 
The tasks of complex reasoning and validation are thus often executed by two independent systems, which creates a technical overhead. OWL reasoners are furthermore not always well maintained \cite{abicht2023,LamElvesaeterRecuerda2015} 
and not all reasoning results are relevant for validation.
Ke et al.~\cite{DBLP24} address the latter problem in their paper.
They carefully consider which facts need to be materialised and combine  this with a slight rewriting. They present an efficient way to perform SHACL validation in presence of OWL~LD~\cite{DBLP:conf/www/GlimmHKP12}. However, their approach requires two distinct steps for each data graph: first the  RE-SHACL system needs to be run, followed by a standard SHACL validator. We consider another ontology language that is not contained in OWL~LD and  aim to go another route; we propose a data independent rewriting. Based solely on the ontology axioms and shapes graph, we produce a new shapes graph.  This shapes graph can then be applied to any input data graph with 
any SHACL validator.

This is illustrated by the following example.
We show a constraint which makes sure that every person has at least one name.\footnote{We omit the prefix declarations for brevity.}
\begin{center}
\begin{minipage}{0.45\linewidth}
\begin{lstlisting}[style=turtle,basicstyle=\ttfamily\scriptsize]
:PersonShape
    a sh:NodeShape ;
    sh:targetClass :Person ;
    sh:property [
        sh:path :name ;
        sh:minCount 1 ;
    ] .
\end{lstlisting}
\end{minipage}
\end{center}
\sloppy
Moreover, we know that every student is a person, 
\lstinline[style=turtle,basicstyle=\ttfamily,frame=none]{:Student rdfs:subClassOf :Person.}. Here, we can add a target declaration 
\lstinline[style=turtle,basicstyle=\ttfamily,frame=none]{:PersonShape sh:targetClass :Student.}
to make sure that the shape \texttt{:PersonShape} also targets all instances of the class \texttt{:Student}. 

This does not only work for  targets but also for constraints. If the ontology, for example, contains a subproperty
\mbox{\lstinline[style=turtle,basicstyle=\ttfamily,frame=none]{:studentName rdfs:subPropertyOf :name.}},
we can replace the IRI
\mbox{\lstinline[style=turtle,basicstyle=\ttfamily,frame=none]{:name}}
(line 5) by 
\lstinline[style=turtle,basicstyle=\ttfamily,frame=none]{[ sh:alternativePath (:name :studentName) ]},
and ask for a \emph{name} or a \emph{student name} instead.
The idea is thus that all constraint violations that could be found by first performing ontology reasoning and then executing
constraint checking can also be found by only performing constraint checking using the rewritten shapes. 
The advantage of this approach is that the results become independent of the reasoning a SHACL validator supports and are therefore well-suited to be shared through the Web. In practical set-ups with fixed ontologies and shapes, but varying data graphs, only one system, the SHACL validator, needs to be maintained.

Note that the idea behind this form of rewriting differs from classical rewriting of SPARQL \cite{sparqlspec} queries based on OWL-QL \cite{DBLP:conf/rweb/Krotzsch12}  as first proposed by Poggi et al. \cite{obda-idea} and realised in many applications \cite{obda-surv}.
When we rewrite SPARQL  based on OWL QL axioms, we modify one single existing query. This does not allow for any recursion beyond SPARQL's property paths. 
A SHACL graph consists of several shapes which may depend on each other and while recursion between different shapes is not yet part of the standard, this extension is planned for the next version \cite{shaclCharter}.

The SHACL working group furthermore plans to add rule support \cite{shacl12-rules} to the specification. Engines supporting this feature will naturally be able to cope with RL using the corresponding rules \cite[Section 4.3]{DBLP:conf/rweb/Krotzsch12}. 
Being expressible in Datalog, these rules could alternatively be directly incorporated into SHACL shapes using the rewriting Pareti et al. provided \cite{shacl-datalog}.
We therefore do not focus on the RL fragment here.

There have been different contributions to ontology-based SHACL rewriting: Savkovic et al. provide a rewriting of positive SHACL constraints based on OWL QL axioms \cite{DBLP:conf/esws/SavkovicKL19}. 
This got extended in Ahmetaj et al.  \cite{ecai2023} to a rewriting for OWL QL and SHACL constraints with a restricted form of negation and recursion. However, their algorithm is best-case exponential and can thus not be used in practice. 
Oudshoorn et al. extended the former for the description logics \elhi\ \cite{dl2024} and Horn-$\mathcal{ALCHIQ}$ \cite{aij2026}, which both subsume OWL QL. They present a rewriting procedure for \elhi\ that is no longer best-case exponential (only worst-case, but given their \exptime-completeness results, this is inevitable) and thereby provide a first step towards an implementation.
However, 
all of these contributions stayed on a theoretical level. 
If we broaden our horizon beyond SHACL, there are more approaches~\cite{Fan:2010:RIC:1862919.1862924,DBLP:journals/ai/KnorrAH11,DBLP:conf/owled/MotikHS07,DBLP:conf/www/MotikHS07} which discuss the theory of integrating constraints and implicit knowledge or OWL. 

In this paper we make the step from theory to practice: 
we present \owlawareshacl, a tool which rewrites SHACL shapes based on OWL ontologies to enable ontology-aware validation.
We extend the existing theory by considering nominals, role chains and qualified existential restrictions on the left-hand side of the ontology axioms. However, we limit the usage of existential restrictions on the right-hand side as these are known to be the main reason for the exponential blow-ups 
\cite{ecai2023,dl2024}. The resulting fragment \owlelm still supports non-trivial axioms and is expressive enough to be used in practical applications \cite{riskman}. Identifying this fragment for which we can extract an implementable algorithm, compared to the theoretically heavy rewriting techniques presented in related work (that are best case exponential size), is our first contribution.
Our second contribution is to adapt existing rewriting techniques to our setting. In particular, we describe a new approach for rewriting targets as well.
We provide an implementation, and perform an extensive evaluation using different OWL reasoners and SHACL validators. 
We compare the execution times between two approaches: (1) we rewrite the shapes based on the ontology and then validate the data graph against the rewritten shapes graph, 
(2) we reason on the data graph using the ontology and then validate the result against the original shapes graph. 
The pure validation times are  similar in both approaches.
This backs up our claim that rewritten shapes can be shared and used in practical applications. 
Links to the implementation source code and benchmarks are provided in the supplementary material statement before the references.

%% file: sections/02-preliminaries.tex
% !TEX root = ../main.tex
\section{Preliminaries}\label{sec:preliminaries}

Before getting to the body of this paper, we define the following notions.

\paragraph{Data Graphs.}
Let $\NC,\NR$ and $\NI$ denote countably infinite, mutually disjoint sets of \emph{concept names} (also known as \emph{class names}), \emph{role names} (or, \emph{property names}), and \emph{individuals} (or, \emph{constants}), respectively. We assume $\{\top,\bot\} \subseteq \NC$.
Let $\roles := \{p,p^- \mid p \in \NR\}$ denote the set of \emph{roles}. In general, we will write $p \in \NR$ and $r \in \roles$, to denote the distinction. For every $p\in \NR$, let $(p^-)^- = p$. 
An \textit{atom} (or, \emph{assertion}) is an expression of the form $A(c)$ or $p(c,c')$, for $A\in \NC$, $p\in \NR$ and $\{c,c'\} \subseteq \NI$. A \emph{data graph} $\A$ is a finite set of atoms.

An \emph{interpretation} is a pair $\I=(\Delta^{\I},\cdot^{\I})$, where $\Delta^{\I}$ is a non-empty set, called the \emph{domain}, and $\cdot^{\I}$ is a function that maps every $A\in \NC$ to a set $A^{\I}\subseteq \Delta^{\I}$, every $p\in \NR$ to a binary relation $p^{\I}\subseteq \Delta^{\I}\times \Delta^{\I}$, and every individual $c\in \NI$ to an element $c^{\I}\in \Delta^{\I}$. Let $(p^-)^\I := \{(c',c) \mid (c,c') \in p^\I\}$.
The \textit{canonical interpretation} $\I_\A$ of a data graph $\A$ is defined by letting $\Delta^{\I_\A}$ contain all individuals
occurring in $\A$, and setting $c^{\I_{\mathcal{A}}} := c$ for
each $c \in \NI$, $A^{\I_\A} := \{ c \mid A(c) \in \A\}$
for each $A \in \NC$ and $r^{\I_\A} := \{(c,c') \mid r(c,c') \in \A\}$ for all $r \in \roles$.
We make the standard name assumption, which means $c^\I = c$ for all interpretations $\I$, and all $c \in \NI$. Note that this enforces a unique name assumption too.

\paragraph{(Non-recursive) SHACL.}
Let $\NS$ be an infinite set of shape names $s$. Based on the work of Corman et al. \cite{DBLP:conf/semweb/CormanRS18}, we say a \textit{(SHACL) constraint} $s \gets \varphi$ is formed from a shape name $s \in \NS$ and a shape expression $\varphi$, defined in the following way
\begin{align*}
  \varphi::=  c\mid s \mid A \mid \neg \varphi \mid
  \varphi \sqcap \varphi
  \mid \;\ggan E.\varphi 
  \mid E = E'
  \mid E \not = E',
\end{align*}
where $c \in N_I$, $s \in N_S$, $A \in N_C$, $n \geq 1$, $p \in \NR$ 
and $E$ and $E'$ regular path expressions, i.e., regular expressions over the language $\roles$, defined as
\begin{align*}
  E ::= r \mid E \cup E \mid E \cdot E \mid E^*.
\end{align*}
We use $\forall E.\varphi$ as a shorthand for $\lnot \!\!\ggae\! E.\lnot \varphi$, and $\varphi \sqcup \varphi'$ for $\lnot (\lnot \varphi \sqcap \lnot \varphi')$. Note that a similar trick does not work for $E = E'$ and $E \not = E'$, as sets of atoms being equal or completely disjoined cannot be simply expressed in terms of each other by using negation.
Let $L_E$ be the language defined by some regular expression $E$, containing words $w \in \Sigma^*$, inductively defined as $L_r := \{r\}$, $L_{E \cup E'} := L_E \cup L_{E'}$, $L_{E \cdot E'} := \{w \cdot w' \in \Sigma^* \mid w \in L_E, w' \in L_{E'}\}$, and $L_{E^*} := \{w^* \in \Sigma^* \mid w \in L_E\}$. We say $E$ is a regular \textit{path} expression, when $\Sigma = \roles$. 
The semantics of regular path expressions is given in terms of the evaluation $E^\I$ over an interpretation $\I$, which is defined as follows. 
A pair $(e,e')$ is contained in $ E^\I$ iff there exists $r_0\cdots r_n \in L_E$ and $\{e_1,\ldots e_n\} \subseteq \Delta^\I$ such that $(e,e_1) \in r_0^\I$, $(e_n,e')\in r_n^\I$ and for all $1 \leq i \leq n-1$, $(e_i,e_{i+1}) \in r_i^\I$.

A \textit{constraint set} $\C$ is a set of SHACL constraints such that for each $s\in \NS$, there exists at most one $s\gets \varphi \in \C$, and such that there are no cyclic dependencies.
The semantics of SHACL is defined in a recursive manner by the function $\I(\cdot)$, given in Figure \ref{fig:semantics}. 
\begin{figure}[t]
  \centering
  \begin{align*}
    \I(c) &:= \{c^{\I}\} \\
    \I(s) &:= \{e \in \varphi^\I \mid s \gets \varphi \in \C\} \\
    \I(A) &:= A^{\I}  \\
    \I(\lnot \varphi) &:= \Delta^\I \setminus \I(\varphi) \\
    \I(\varphi \sqcap \varphi') &:= \I(\varphi)\cap \I(\varphi') \\
    %\I(\varphi\lor \varphi') &:= \I(\varphi)\cup \I(\varphi') \\
    \I(\exists_{\geq n}E.\varphi) &:= \{e \in \Delta^\I \mid |\{e' \in \Delta^\I \mid (e,e') \in E^\I \land e' \in \I(\varphi)\}| \geq n\} \\
    \I( E = E') &:= \{e \in \Delta^\I \mid \{\exists e' \in \Delta^\I\mid (e,e') \in E^\I\} = \{e' \in \Delta^\I \mid (e,e') \in E'^\I\}\}\\
    \I(E \not = E') &:= \{e \in \Delta^\I \mid \{\exists e' \in \Delta^\I\mid (e,e') \in E^\I\} \cap \{e' \in \Delta^\I \mid (e,e') \in E'^\I\} = \emptyset\}
  \end{align*}
  \caption{Evaluating shape expressions}
  \label{fig:semantics}
\end{figure}
Given an interpretation $\I$ and a shape assignment $S$, we say a node $c \in \NI$ \textit{validates} a shape expression $\varphi$, when $c \in \I(\varphi)$. 
Furthermore, let $\G$ be a set of targets of the form $s(X)$, for
\[ X \eqbnf c \mid A \mid \exists r.\top  \]
where $c \in \NI$, $A \in \NC$ and $r \in \roles$.
A pair $(\C,\G)$ consisting of a constraint set and set of targets is called a \textit{shapes graph}. Given an interpretation $\I$, we say $\I$ \textit{validates} $(\C,\G)$ if for all $s(c) \in \G$, we find $c$ validates $s$, and for all $s(X) \in \G$, all nodes in $X^\I$ validate $s$. Considering readability, we will write $\A$ validates $(\C,\G)$, for any set of atoms $\A$, in which case the canonical interpretation $\I_\A$ is intended.

\section{Ontology Language \owlelm}
\label{sec:owlelm}

For the ontology axioms, we introduce the language OWL EL$^-$, a fragment of OWL EL. This logic corresponds to the description logic \elpp \cite{DBLP:conf/ijcai/BaaderBL05,DBLP:conf/owled/BaaderLB08}, but is disallowing the usage of existential restrictions on the right-hand side. An example of an ontology expressible in this fragment in the Riskman ontology~\cite{riskman}.

Let an \owlelm TBox $\T$ be a set of axioms of the form $r \isa p$, for $r \in \{p', p' \circ p \mid p' \in \NR\}$, or $C \isa B$, where
\begin{align*}
   C \eqbnf \top \mid A \mid c \mid \ran(p) \mid C \sqcap C \mid \exists p.C 
\end{align*}
where $\{A,B\} \subseteq \NC$, $c \in \NI$ and $\{p,p'\} \subseteq \NR$. Furthermore, we use $\dom{p}$ as a shorthand for $\exists p.\top$. Any expression that may occur on the left-hand side of an \owlelm axiom is referred to as an \owlelm \textit{concept}.

\iffalse
That is, we consider every axiom to be of one of the following inclusion axioms, in normal form:
\begin{align*}
  C_1                          & \isa B & \ran(p)    & \isa B  \\
  C_1 \sqcap \ldots \sqcap C_n & \isa B & p \circ p' & \isa p' \\
                               &        & p          & \isa p'
\end{align*}

  \begin{align}
    \exists p. C                 & \isa B                            \\
    A_1                          & \isa \exists p. a                 \\
    A_1 \sqcap \ldots \sqcap A_n & \isa B                            \\
    \ran(r)                      & \isa B                            \\
    r \circ r'                   & \isa r' \label{roleconcatenation} \\
    r                            & \isa r'
  \end{align}
\fi
Note that, to avoid intractability of this logic, we reduced the syntax of axioms of the form $p' \circ p \isa p''$ to $p \circ p \isa p$. This is a slightly weaker form of the syntactic restriction proposed by Baader et al. \cite{DBLP:conf/owled/BaaderLB08}, which ensures tractability. Note that transitivity axioms may still be expressed using the form $p \circ p \isa p$. 

The semantics of \owlelm is, as is usual, defined in terms of interpretations $\I$: an axiom $C \isa D$ is satisfied whenever $C^\I \subseteq D^\I$. To this end, the interpretation function is extended in the following way: 
$\top^\I := \Delta^\I$, $(C \sqcap C')^\I := C^\I \cap C'^\I$, $(\exists p.C)^\I := \{c \mid (c,c') \in p^\I \land c' \in C^\I\}$, $(\ran(p))^\I := \{c \mid (c',c) \in p^\I\}$, and $(p' \circ p)^\I := \{(c,c') \mid (c,d) \in p'^\I, (d,c') \in p^\I\}$. 
In case all axioms in $\T$ are satisfied in $\I$, we say $\I$ is a \textit{model} of $\T$. As is standard, we define subformulas in the following recursive way: $\sub(\T) := \{\sub(X),sub(B) \mid X \isa B \in \T\}$, and $\sub(C \sqcap C') := \{C \sqcap C',\sub(C),sub(C')\}$, $\sub(\exists p.C) := \{\exists p.C, \sub(C)\}$ and $\sub(Y) := \{Y\}$ for $Y \in \NC \cup \NI \cup \{\top,\ran(p)\mid p \in \NR\}$. 

\paragraph{Syntactic Restrictions.}
To ensure that counting to at least two on complex path expressions is restricted to simple roles, we assume the following restriction on $\C$ and $\T$ to be honoured: if $\ggan p.\varphi$ appears somewhere in $\C$, there do not exist $p' \circ p \isa p \in \T$ or $p' \isa p \in \T$.

Moreover, in the rewriting itself we are considering \textit{acyclic} TBoxes: for each $\T$, we assume that there exists no $\{C_1 \isa D_1,\ldots,C_n \isa D_n\} \subseteq \T$ such that for each pair $D_i$ and $C_{i+1}$, and the pair $D_n, C_1$, there is a concept or role name in the intersection of the syntax of the pair. 

We note that both cyclic TBoxes as recursive SHACL do not break our theory, nor the correctness of our rewriting. However, to capture the cyclicity, we would produce (stratified) recursive SHACL constraints, that would need to be interpreted under a least fixed point semantics.\footnote{Note that the stable-model and well-founded semantics would also suffice, as they coincide with the least fixed point semantics for stratified constraints.} To the best of our knowledge, the only (proof-of-concept) implementation that handles recursion in a principled way is shaWell \cite{DBLP:conf/kr/OkulmusS24}. That is, to allow for a more extensive evaluation, we stick to non-recursive SHACL and acylic TBoxes in this paper.

%% file: sections/03-warm-up.tex
% !TEX root = ../main.tex
\section{Combining SHACL and OWL}\label{sec:warmUp}
We aim to  provide a  feasible method to combine OWL reasoning and SHACL validation in practical applications.
To this end, we first need to clarify which semantics we consider. 
%Before detailing our method, we clarify which semantics we consider. 
%In this paper we aim 
Several papers investigated the theory of this combination \cite{ecai2023,dl2024,aij2026,DBLP:conf/esws/SavkovicKL19}, and all of them have in common that SHACL constraints are evaluated over some minimised canonical model of the original data graph and OWL constraints. These papers mention multiple ways to define such a minimal model: minimisation after Skolemisation, or using the core operation on any universal model. %This is where the first advantage of disallowing existential restrictions on the right shows up: 
As we disallow existential restrictions on the right hand side of an axiom, we do not face the problem these definitions aim to solve:
for all new information we possibly derive using axioms, we always know which individuals are affected. Building a unique universal model that is minimal is thus straightforward.

\begin{df}\label{def:at}
     Given a set of \owlelm axioms $\T$ and a data graph $\A$, let $\A_\T$ be the smallest set such that $\A \subseteq \A_\T$ and moreover:
     \begin{itemize}
         \item for each $C \isa B \in \T$, for $C$ an \owlelm concept if $c \in C^{\I}$, then $B(c) \in \A_\T$;
         \item for each $r \isa p \in \T$, for $r \in \{p',p'\circ p \mid p' \in \NR\}$, if $(c,c') \in r^{\I}$, then $p(c,c') \in \A_\T$,
     \end{itemize}
     where $\I$ is a shorthand for $\I_{\A_\T}$. 
\end{df}

The following is immediate.

\begin{pr}
    Given a set of \owlelm axioms $\T$ and a data graph $\A$, we find that the canonical interpretation of $\A_\T$ is a model of $(\T,\A)$. Moreover, there does not exist a smaller model of $(\T,\A)$: for each model $\J$ of $(\T,\A)$, we find that $A^\I \subseteq A^\J$ and $p^\I \subseteq p^\J$, for all $A \in \NC$ and $p \in \NR$.
\end{pr}

\iffalse
\begin{df}\label{def:at}%[Minimal model $\A_\T$]\label{def:at}
     Given a set of \owlelm axioms $\T$, define an immediate consequence operator $T_\T$ that maps a set of atoms $X$ to a set of atoms as follows:
    \begin{align*}
        T_\T(X) := X \cup 
         &\{B(c) \mid \{p(c,c'),C(c')\} \subseteq X, \exists p.C \eqhaak B \in \T\} \\
         %\cup &\{p(c,a) \mid A_1(c) \in X, A_1 \isa \exists p.a \in \T\} \\
       \cup &\{B(c) \mid \{A_1(c),\ldots,A_n(c)\} \subseteq X, A_1 \sqcap \ldots \sqcap A_n \eqhaak B \in \T\} \\
    \cup &\{B(c) \mid p(c',c) \in X, \ran(p) \isa B \in \T\} \\
    \cup &\{p(c,c') \mid p'(c,c') \in X, p' \eqhaak p \in \T\} \\
    \cup &\{p(c,c') \mid \{p'(c,d),p(d,c')\} \subseteq X, p' \circ p \eqhaak p \in \T\}.
    \end{align*} 
    Given any data graph $\A$, let $\A_\T$ be the smallest set containing $\A$ such that $T_\T(\A_\T) = \A_\T$. 
\end{df}
\fi

This brings us to the semantics of SHACL in presence of \owlelm axioms.

\begin{df}
    Given a data graph $\A$, a set of \owlelm axioms $\T$ and a shapes graph $(\C,\G)$. We say $(\T,\A)$ validates $(\C,\G)$ whenever $\A_\T$ validates $(\C,\G)$.
\end{df}

The rest of this paper focuses on how we can replace this two-stage method by a more direct one. We want to build a set of SHACL constraints, based on a shapes graph $(\C,\G)$ and a set of axioms $\T$, that can directly be validated against any data graph. % That is, can we internalise the OWL reasoning in SHACL constraints? 
We will show the following:

\begin{theorem}
    Given a shapes graph $(\C,\G)$ and an \owlelm TBox $\T$, there exists a shapes graph $(\C_\T,\G_\T)$ such that for every data graph $\A$, 
$$
\A_\T \text{ validates } (\C,\G) \text{ iff } \A \text{ validates } (\C_\T,\G_\T). 
$$
\end{theorem}

The main advantage of updating the shapes graph in this way is that for every data graph, and every update of any of these data graphs, it suffices to use $(\C_\T,\G_\T)$ -- which only needs to be computed once -- when testing validation. 

The rewriting approach consists of two key phases:
\begin{itemize}
  \item \define{constraint rewriting} -- integration of the axioms in $\TBox$ into the input constraints $\C$.
  \item \define{target rewriting} -- using the constraint rewriting, fine-tuning targeting such that also targets implied by $\T$ are addressed.
\end{itemize}

To understand the distinction between the two aspects, consider the following example.

\begin{example}\label{ex:two-types-of-rewriting}
  Let $\TBox=\set{\exists p.A \sqsubseteq B, p\sqsubseteq q}$, $\ABox=\set{A(b), p(a,b)}$, $\C=\set{s\gets \exists q.\top}$, $\G=\set{s(B)}$ be the sets of ontology axioms, assertions, shapes and targets, respectively. In this set-up, we get $\A_\T = \A \cup \set{B(a), q(a,b)}$. 
First, we want to internalise in the shapes graph that nodes like $a$ conforming to $\exists p.A$ should also be considered a target for $s$. This part is addressed in the target rewriting. Second, since $q(a,b) \in \A_\T \setminus \A$, we would like to update the constraint set to $s \gets \exists (p \cup q).\top$, such that we may indeed conclude that $a$ validates $s$. 
\end{example}

In the following section, we go into more detail of these rewriting techniques.

\iffalse
\begin{itemize}
  \item realisation of $\ABox$ against $\TBox$, denoted:
  \[
    \real{\ABox}{\TBox}\eqdef \set{\alpha\guard\ABox\cup\TBox\entails\alpha}
  \]
  \pg{This, but restricted to only individuals named in $\ABox\cup\TBox$}
  \item validation result of $\ABox$ against set of shapes $\C$ and targets $\G$.
  \[
    \val{\ABox}{\tuple{\C,\G}} \mapsto \{\top, \bot\}
  \] 
\end{itemize}
Then, given $\TBox$, $\C$ and $\G$ our procedure $\rew{\TBox}{\tuple{\C,\G}}=\tuple{\C', \G'}$ our main claim is: for any ABox $\ABox$:
\[
  \val{\real{\ABox}{\TBox}}{\tuple{\C, \G}} = \val{\ABox}{\rew{\TBox}{\tuple{\C,\G}}}
\] 
\fi

%% file: sections/04-rewriting.tex
\section{Rewriting Techniques}
\label{sec:rewriting}

The main idea of the rewriting is to capture all reasoning of the axioms in SHACL constraints. The most direct translation of the axioms can be found in the $\shape(X,\TBox)$-part: here, it is ensured that if a node is labelled with $s_C$ for some concept $C$, then $C$ must be derivable by $\T$ for that specific node. That is, let $\shape(X,\TBox)$ be defined as

\begin{align*}
         \shape(X,\TBox) & \eqdef
        \begin{cases}
            s_a                 \gets a               & \text{if } X = a                             \\
            s_{\exists p.C}     \gets \exists p.s_C   & \text{if } X = \exists p.C                   \\
            s_{\exists p.\top} \gets \exists p.\top & \text{if } X = \dom{p}                 \\
            s_{\exists p^{-}.\top} \gets \exists p^-.\top & \text{if } X = \rang{p}                 \\
            s_{C_1\sqcap\ldots\sqcap C_n} \gets s_{C_1}\sqcap \ldots \sqcap s_{C_n}
                    & \text{if } X = C_1\sqcap\ldots\sqcap C_n     \\
            s_B \gets B                               & \text{if } X = B, B \in \NC                           \\[-4pt]
            \phantom{s_B \gets}\;\;\sqcup\;
            \mathrlap{\bigsqcup_{\set{p \guard \rang{r}\sqsubseteq B \in \TBox}} s_{\exists p^-.\top}} \\
            \phantom{s_B \gets}\;\;\sqcup\;
            \mathrlap{\bigsqcup_{\set{p \guard \dom{r}\sqsubseteq B \in \TBox}} s_{\exists p.\top}}    \\
            \phantom{s_B \gets}\;\;\sqcup\;
            \bigsqcup_{\set{C \guard C\sqsubseteq B \in \TBox}} s_C
        \end{cases}
\end{align*}

A standard argument based on the induction on the construction of $C$ suffices to conclude the following.
\begin{lm}
    Given an \owlelm TBox $\T$, let $\C := \bigcup_{X \in \sub(\T)} \shape(X,\T)$, then for all $C \in \sub(\T)$ and $c \in \NI$, we have $c \in \I_\A(s_C)$ iff $c \in C^{\I_{\A_\T}}$.
\end{lm}

This is then used in multiple ways.
First of all, to update the targets of shapes according to $\T$, we use the so-called \textit{bridge}-shapes: if $s(X) \in \G$, we add $s^{\Join} \gets \neg s_X \sqcup s$, for $s^{\Join} $ a fresh shape name that universally targets all nodes but constrains only those satisfying $X$, ensured by an implication encoded as a disjunction. 
In practice, we use the class, property, and individual declarations to perform the universal targeting.\footnote{Note that tools like for example Protegé \cite{protege} automatically add such declarations. % during ontology creation. 
} If the ontology contains, e.g., a triple \mbox{\lstinline[style=turtle,basicstyle=\ttfamily,frame=none]{:r a owl:ObjectProperty.}}, we add target declarations for subjects and objects of \texttt{:r} using \texttt{sh:targetSubjectOf} and \texttt{sh:targetObjectOf}. 
%That is, we target all nodes, collected by using the class, property, and individual declarations, 
We then filter out the nodes which can be labelled by $s_X$. For those nodes, we ensure that $s$ is validated, as is required.
Thus, let the set of targets be updated to
\begin{align*}
%\bridge(s, X)        & \eqdef s^{\Join} \gets \neg(s_X) \sqcup s\\
    \G_\T                  & = \G \cup \set{ s^{\Join}(\top) \guard s(X) \in \G}.
\end{align*}

\newcommand{\role}{\mathit{role}}

Next, to take care about the role inclusions, we first collect the relevant role dependencies. Here, $\to_{\role}$ and $\to_{\chains} $ are two relation given by $p' \to_\role p$ iff $p' \isa p \in \T$, respectively $p' \to_{\chains} p$ iff $p' \circ p \isa p \in \T$. With $*$ and $+$ we denote the reflexive-transitive and transitive closure of relations. That is, let
\begin{align*}
       \Base(\TBox,p)       & \eqdef \set{ p' \guard p' \to_\role^{*} p}                                  \\
        \Dep(\TBox,p)        & \eqdef \set{ p' \guard p' \to^{+}_{\chains} p }\\
        \PathSigma(\TBox,p)  & \eqdef \begin{cases}
            \bigcup_{t\,\in\,\Dep(\TBox,p)} \Base(\TBox,t)
         &\text{if } p \circ p \isa p \in \T \\
        \bigcup_{t\,\in\,\Dep(\TBox,p)} \Base(\TBox,t)
        \setminus \set{p} &\text{otherwise. }
        \end{cases}
\end{align*}
With these dependencies defined, we can now provide the expressions that will be substituted for occurrences of $p$ respectively $p^-$.
\begin{align*}
        L^{+}(\TBox,p)       & \eqdef (p_1 \cup \cdots \cup p_k)^{*} \cdot (p'_1 \cup \cdots \cup p'_n)\\
        L^{-}(\TBox,p)       & \eqdef (p'^{-}_1 \cup \cdots \cup p'^{-}_n) \cdot (p^{-}_1 \cup \cdots \cup p^{-}_k)^{*}                                            
\end{align*}
Here, $\{p_1,\ldots,p_k\} = \PathSigma(\TBox,p)$ and $\{p_1',\ldots,p_n'\} = \Base(\TBox,p) $.
To illustrate these auxiliary definitions for roles consider the following example.

\begin{example}
        Let $\set{p_0 \sqsubseteq p_1, p_1 \circ p_3 \sqsubseteq p_3, p_2 \sqsubseteq p_3}\subseteq \TBox$, and $p_3$ and $p_3^-$ appearing in shapes.
        Then %$\mathit{chains}(\TBox, r_3) = \set{r_1}$, 
        $\Base(\TBox, p_3) = \set{p_2, p_3}$, $\Dep(\TBox, p_3) = \set{p_1}$,
        $\PathSigma(\TBox, p_3) = \set{p_0, p_1}$, and $p_3$ will be substituted by $(p_0 \cup p_1)^* \cdot (p_2 \cup p_3)$,
        whereas $p_3^-$ by $(p_2^- \cup p_3^-) \cdot (p_0^- \cup p_1^-)^*$.
        Since $p_3 \circ p_3 \sqsubseteq p_3 \notin \TBox$, $p_3$ is excluded from
        $\PathSigma(\TBox, p_3)$ to prevent admitting repeated $p_3$-steps in the prefix;
        had $p_3$ been transitive, the path would expand to $(p_0 \cup p_1 \cup p_3)^* \cdot (p_2 \cup p_3)$ instead.
\end{example}

In the rewriting we need to substitute certain roles or concept names by fresh objects, that is, let $\varphi\bigl[ x \mapsto y(x) \bigr]_{x\in X}$ and $\C\bigl[ x \mapsto y(x) \bigr]_{x\in X}$ be the notation indicating that for each $x \in X$, every occurrence of $x$ in $\varphi$ resp.\ $\C$ is replaced by $y(x)$, where for all $p \in \NR$, we consider $p^-$ a symbol in itself, not containing $p$.

\begin{lm}
   Let $\T$ be any \owlelm TBox only containing axioms of the form $r \isa p$, and $\varphi$ any shape expression, then we find that for each data graph $\A$, we have $c \in \I_{\A_{\T}}(\varphi)$ iff $c \in \I_\A(\varphi\bigl[\,p \mapsto \LPos(\TBox,p),\;
        p^{-} \mapsto \LNeg(\TBox,p)\,\bigr]_{p\,\in\,\NR})$ for all $c \in \NI$.
\end{lm}

The main idea why this substitution suffices is that for every axiom of the form $p \isa p'$, $p'$ is replaced by $p \cup p'$, whereas for axioms of the form $p \circ p' \isa p'$, $p'$ is replaced by $p^* \cdot p'$. The sets $\Sigma(\T,p)$ and $\Base(\T,p)$ solely collect the whole set of dependent $p$'s, and do the replacements simultaneously to avoid termination issues. For the inverse substitution, dictated by $L^-$, exactly the same is happening, but in reverse.

\iffalse
\begin{lm}
    Some lemma on that the above does what it needs to do for targets
\end{lm}
\fi

For the constraint rewriting, the main idea is to substitute all concept names $B$ appearing in the constraint set $\C$ by the shape name $s_B$, which, as mentioned before, collects exactly all nodes for which $B$ can be derived in $\T$.%: $\C\bigl[\,B \mapsto S_B\,\bigr]_{{B\,\in\,\NC}}$.

That is, the final rewritten set of shapes, $\C_\T$ is given by
    \begin{align*}
        \allShapes(\TBox,\C,\G) & \eqdef \C\bigl[\,B \mapsto S_B\,\bigr]_{{B\,\in\,\NC}} \cup \set{s^{\Join} \gets \neg s_X \sqcup s \guard s(X) \in \G} \\
         & \phantom{\eqdef \C_{\mathrm{norm}}\;} \cup \set{\shape(X,\TBox) \guard X\in \sub(\T) \cup \{X \mid s(X) \in \G}                                                       \\
        \C_\T                  & \eqdef \allShapes(\TBox,\C,\G)
        \bigl[\,p \mapsto \LPos(\TBox,p),\;
        p^{-} \mapsto \LNeg(\TBox,p)\,\bigr]_{p\,\in\,\NR}.                                                      
    \end{align*}

In a concrete setting, this produces the following constraints.

\begin{example}\label{ex:running-example-declarative}
    Consider the following TBox and shapes graph
\begin{align*}
        \TBox & = \set{ A \sqsubseteq B,\ \exists p.B \sqsubseteq C,\ \dom{p} \sqsubseteq A,\ \rang{p} \sqsubseteq B,\ p_0 \sqsubseteq p,\ p_1 \circ p \sqsubseteq p,\ p_0 \sqsubseteq p_1 } \\
        \C    & = \set{ s_1 \gets \exists p.B,\quad s_2 \gets \exists p^-.A } \quad
        \G = \set{ s_1(A),\quad s_2(\exists p^-.\top) },
    \end{align*}
    following the rewriting as described above, this produces the following rewritten shapes graph
    \begin{align*}
        \C_\T & = \left\{\begin{array}{l}
                           s_1 \gets \exists\bigl((p_0 \cup p_1)^* \cdot (p_0 \cup p)\bigr).s_B, \quad
                           s_2 \gets \exists\bigl((p_0^- \cup p^-) \cdot (p_0^- \cup p_1^-)^*\bigr).s_A                   \\[4pt]
                           s_A \gets A \sqcup s_{\exists p.\top}, \quad
                           s_B \gets B \sqcup s_A \sqcup s_{\exists p^-.\top}, \quad
                           s_C \gets C \sqcup s_{\exists p.B}                                                               \\[4pt]
                           s_{\exists p.\top} \gets \exists\bigl((p_0 \cup p_1)^* \cdot (p_0 \cup p)\bigr).\top           \\
                           s_{\exists p^-.\top} \gets \exists\bigl((p_0^- \cup p^-) \cdot (p_0^- \cup p_1^-)^*\bigr).\top \\
                           s_{\exists p.B} \gets \exists\bigl((p_0 \cup p_1)^* \cdot (p_0 \cup p)\bigr).s_B               \\[4pt]
                           s_1^{\Join} \gets \neg s_A \sqcup s_1, \quad
                           s_2^{\Join} \gets \neg s_{\exists p^-.\top} \sqcup s_2
                       \end{array}\right\} \\[6pt]
        \G_\T & = \G \cup \set{ s_1^{\Join}(\top),\quad s_2^{\Join}(\top) }.
    \end{align*}
\end{example}

\iffalse
\begin{cor}
    Given a shapes graph $(\C,\G)$ and an \owlelm TBox $\T$, there exists a shapes graph $(\C_\T,\G_\T)$ such that for every data graph $\A$, 
$$
\A_\T \text{ validates } (\C,\G) \text{ iff } \A \text{ validates } (\C_\T,\G_\T). 
$$
\end{cor}
\fi

\begin{prooftheorem}
    The correctness of our rewriting follows from combining the intuitions of Lemmata 1 and 2 with the idea that for each $s \gets \varphi \in \C$, with $s(X) \in \T$, we find that for all $c \in X^{\I_{\A_\T}}$ that $c$ validates $s^{\Join}$ in $\A_\T$ iff $c$ validates $s$ in $\A_\T$, and for all $c \not\in X^{\I_{\A_\T}}$, $c$ validates $s^{\Join}$ in $\A_\T$ anyway, independent of whether $c$ validates $s$ in $\A_\T$ or not. \qed
\end{prooftheorem}

Note that our rewriting is data independent, which means that in size of the data graph, the size of our rewriting is a constant. In size of the in general much smaller TBox, the amount of newly introduced shapes is linear. The size of rewritten constraints itself is also limited: the only blow-up in size may be caused by replacing roles by a regular expression (which size only depends on the amount of role hierarchies in the TBox).

%% file: sections/08-implementation.tex
% !TEX root = ../main.tex
\section{Implementation and Evaluation}\label{sec:implementationandevaluation}

\begin{figure}[ht]
    \centering
    \makebox[\textwidth][c]{\includegraphics[width=\textwidth]{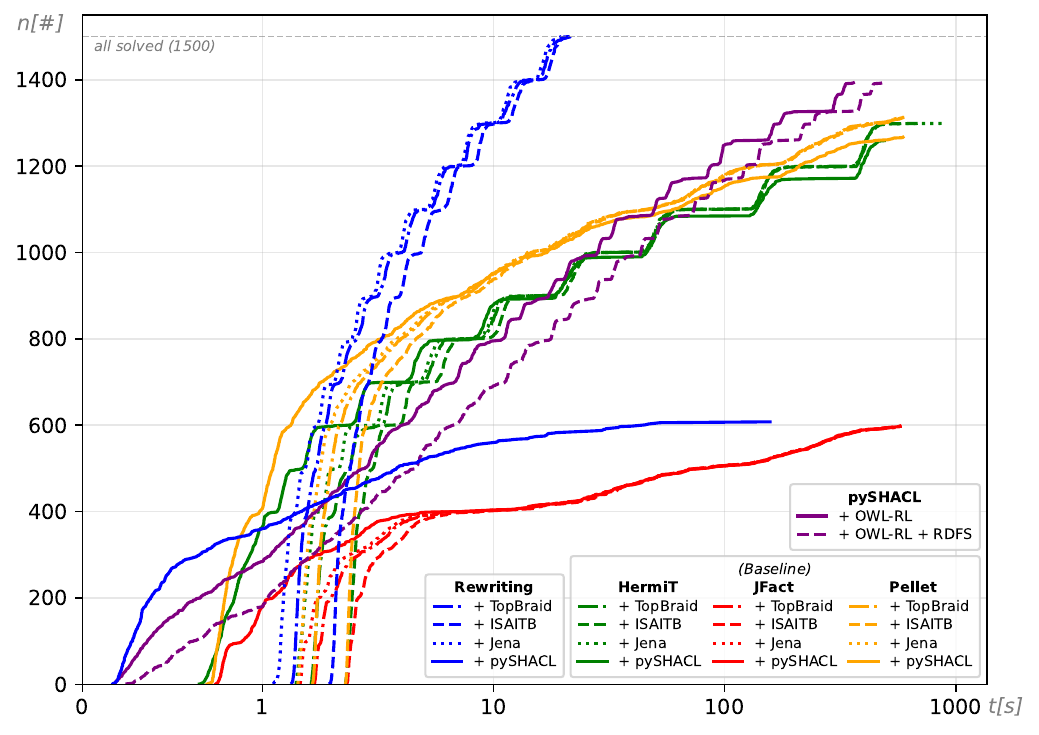}}
    \caption{Cactus plot comparing successful instances across evaluated approaches. The x-axis shows total execution time per instance in seconds (symlog scale: linear below 1s, logarithmic above). The y-axis shows the cumulative number of instances solved within that time; lines reaching higher values indicate better coverage, lines further left indicate faster execution.
    Colors denote the approach: blue -- rewriting, green/red/orange -- baseline (\hermit/\jfact/\pellet), purple -- \pyshacl{} with inference. Line styles distinguish validators (\topbraid: dash-dot, \isaitb: dashed, \jena: dotted, \pyshacl: solid).
    Rewriting (blue) achieves complete coverage of all 1500~instances in under 10s. Among the baseline methods, \hermit{} and \pellet{} reach up to 1300~instances and require up to several hundred seconds; \jfact{} stalls at around 600~instances. Both \pyshacl{} inference modes reach around 1400~instances 
    and are considerably slower than the rewriting approach.
    }
    \label{fig:cactus-plot}
\end{figure}

\begin{table}[h]
    \centering
    \input{sections/evaluation/table}
    \vspace{10pt}
    \caption{Results per approach and validator. T.R./Av.R. -- total/average reasoning or rewriting time; T.V./Av.V. -- total/average validation time; V.E. -- validator errors; T/O: timeouts; 
    Success -- successfully solved instances and success rate. For \pyshacl{} with inference, reasoning and validation are a single step, so only validation statistics are reported.}
    \label{tab:results}
\end{table}

We implemented \owlawareshacl\footnote{Available at \url{https://github.com/gorczyca/owl-aware-shacl}}, a JavaScript tool realising the rewriting shown in the previous section, using the rdflib.js library (v2.2.37) for RDF parsing, graph manipulation, and Turtle serialisation. Given an \owlelm TBox $\TBox$ and SHACL shapes $\tuple{\C,\G}$, it produces rewritten shapes $\tuple{\C',\G'}$ that encode the terminological knowledge directly within the constraints, eliminating the need for a reasoner while providing the same validation results. To verify its correctness and evaluate its performance we designed and carried out a set of experiments and comparisons with related systems, that will be described in this section. 

\paragraph*{Related Systems. }
\owlawareshacl offers an alternative to the baseline method, in which the datagraph is first materialised against the ontology, completing all the implicit class and role assertions. For the inference phase in the baseline method we used three OWL 2 reasoners: \pellet~\cite{pellet} (v2.4.0), \hermit~\cite{hermit} (v1.4.3.517), and \jfact~\cite{jfact} (a Java port of FaCT++, v5.0.3), all integrated via the OWL API~\cite{owlapi} (v5.1.20) through a custom JAR wrapper that returns all inferred class and object property assertions.
We also considered \elk~\cite{elk} (v0.6.0) and \structuredreasoner (the built-in OWL API reasoner), but excluded them as they do not support ABox completion through the OWL API, being primarily designed for other reasoning tasks such as classification.
In the second phase, SHACL validation is carried out. For the \owlawareshacl{} approach this means verifying the initial datagraph against the rewritten constraints; for the baseline method it means validating the materialised datagraph against the original shapes. For validation we used four SHACL validators: \jena~\cite{jena} (v4.10.0), \topbraid~\cite{topbraid-shacl} (v1.4.3), \isaitb~\cite{isaitb-shacl} (v1.10.0), and \pyshacl~\cite{pyshacl} (v0.30.1). We additionally tested \pyshacl's built-in inference feature, which accepts an input ontology alongside the data and supports RDFS, OWL RL, or combined reasoning modes. Since  OWL RL subsumes \owlelm,
we include this as a third evaluated approach. 
In summary, we evaluate the following approaches:
\begin{itemize}
    \item \textit{baseline}: materialisation with an OWL reasoner followed by SHACL validation (3 reasoners $\times$ 4 validators = 12 combinations),
    \item \textit{rewriting}: rewriting with \owlawareshacl{} followed by SHACL validation (4 combinations, one per validator),
    \item \pyshacl \textit{with inference}, using OWL RL or OWL RL+RDFS reasoning (2 combinations).
\end{itemize}

\paragraph*{Benchmarks and Experiments Setup. }
We generated 1500 benchmarks across 15 difficulty levels (100 per level), each consisting of an \owlelm TBox, a set of SHACL shapes, and a data graph. Difficulty scales with ontology size and complexity, varying the number of classes, roles, GCIs, role chains, role inclusions, and shape complexity. All benchmarks are acyclic; while our algorithm would still terminate on cyclic GCIs 
yielding recursive rewritings, current SHACL validators do not support recursive shapes. We therefore restricted evaluation to non-recursive TBoxes -- an assumption that can be dropped once such validators become available. 
Full pseudocodes of the generation algorithms, benchmark statistics, triple counts for benchmarks, rewritings and materialised data graphs as well as and Turtle encodings of the input and output from~\Cref{ex:running-example-declarative} are provided in the appendix. We note that due to the random generation process, the vast majority of instances are negative (i.e., the datagraph does not conform), with positive instances appearing almost exclusively at the lowest difficulty level. Approaches were given 1200s in total per each benchmark, split evenly between the reasoning/rewriting and validation phase, and \pyshacl+inference approach was given full 1200s for its singular task. All experiments were run on an HPC cluster, with each job allocated 64\,GB of memory.

\paragraph*{Results Analysis. } The results of our experiments are presented in~\Cref{fig:cactus-plot} and~\Cref{tab:results}.  The first big column of the table displays the overall time (T.R) and the average time (Av.R.) of the rewriting process (first row) and of the reasoning performed by the different engines (rows 2--4). For all our test cases together the rewriting only took a little bit longer than one hour, while the fastest reasoner, Pellet, needed almost 44 hours to perform a materialisation. 
The reason for this difference partly lies in the complexity of the logics supported by the different systems and therefore has to be treated with some care: while our approach only supports  the  \owlelm fragment, the reasoners we employed are OWL DL reasoners, which need to support more complex reasoning. As all ontologies in our benchmark fall under \owlelm, we expect the results of the reasoners to be the same an \owlelm reasoner, if it existed, would produce. The numbers however show that rewriting can be done in a short amount of time (2.53s on average). Note that the rewriting is furthermore independent of the data graph and only needs to be executed once for each combination of ontology and shapes, while in the reasoning set-up each new data graph requires  a new reasoning run.\footnote{If only one datagraph is updated, incremental reasoners could be used.}

In the second big column of the table, we collected the number of validation errors (V.E), the total time (T.V) and the average time (Av.V) per validation engine. Note, that the number of validation errors is significantly higher for \pyshacl than for all other engines, in particular in combination with the rewriting. The reason is that the version of \pyshacl we used only supports shapes with a depth of up to 30, that is, it does not allow deeply nested shapes.\footnote{pySHACL allows the user to change the maximal depth for the shapes using the option \texttt{--max-depth} \cite{pyshacl}, but this feature was broken and only got fixed with version v0.40.0 \url{https://github.com/RDFLib/pySHACL/issues/315}.}
Such kind of nesting is rather unlikely to occur if shapes are created manually, but it is the expected output of our rewriting. If we take, for example, shape $s_1\in \mathcal{C}$ from Example~\ref{ex:running-example-declarative}, and compare it to its rewritten version $s_1\in\mathcal{C}_\TBox$, we see that the rewriting added a dependency on $s_B$ which in turn depends on $s_A$ and $s_{\exists r^-.\top}$ of which the former again depends on  $s_{\exists r.\top}$. We thus create a chain of four  dependent shapes from a single shape which had no connection to any other.\footnote{Note that the example does not take SHACL's property shapes into account as these are not represented as shapes in the logical representation. The turtle version of the translated shape can be found in the appendix.} 

For the other validation engines the average times for the rewritten shapes are %about 40\% 
higher than the validation times of the original shape on the materialised graph. The main reason for 
that is that this number reflects the average of all \emph{successful} validation runs.
%\footnote{We provide a table only comparing the numbers on instances which could be successfully validated three approaches.} 
Especially for the more complex ontologies the materialisation performed by the reasoners often failed which make these numbers difficult to compare. 
The numbers for the cases which could successfully evaluated by (almost) all approaches can be found in the appendix.
The last column of the table reflects the failures. 
We display 
the number of successful runs of (reasoning or rewriting)+validation run (Success) and the number of runs which resulted in a time out (T/O). Only the rewriting approach was successful in all cases. The timeouts for the other approaches were all caused by the reasoners which were not able to perform materialisation in less than 600s for complex ontologies. With \jfact, we could only cover around 40\% of the cases while the success rate of \hermit and \pellet is around 87\%. In the last row of the table, we also display the reasoning and validation time for \pyshacl enabled with entailment. With this approach, we could cover around 93\% of all cases.

In order to better compare all approaches to each other, we display a cactus plot in Figure \ref{fig:cactus-plot}. On the y-axis we count the number of test cases which were each solved in less than the amount of time displayed on the x-axis.
The timings shown are the combined times for reasoning/rewriting and validation. We, observe that around 400 test cases could each be solved in 1 second or less with the reasoning+validation approach using \pellet and \pyshacl (solid yellow line). 
This approach is the most successful for this short amount of time. However, for an execution time of 4s, the rewriting approach using \topbraid, \isaitb and \jena outperforms all other approaches and in less than 25s it solves all instances of the benchmark. We see that with the known exception of \pyshacl, this result does not depend on the validator. This strengthens our claim that the rewritten shapes can be easily exchanged and used by different parties.

%% file: sections/evaluation/table.tex
\begin{tabular}{l@{\hspace{3pt}}r@{\hspace{3pt}}r@{\hspace{8pt}}|@{\hspace{8pt}}l@{\hspace{6pt}}r@{\hspace{6pt}}r@{\hspace{3pt}}r@{\hspace{3pt}}|@{\hspace{5pt}}r@{\hspace{6pt}}r@{\hspace{2pt}}r@{}}
\toprule
Approach&T.R.&Av.R.&Validator &V.E.&T.V.&Av.V.&T/O&\multicolumn{2}{c}{Success (\%)}\\
\midrule
\multirow{4}{*}{Rewriting}&\multirow{4}{*}{1.05h}&\multirow{4}{*}{2.53s}
 &
 \isaitb &0 &1.11h&2.66s&0 &1500&(100.0\%)\\
 &&&\jena &0 &0.71h&1.71s&0 &1500&(100.0\%)\\
 &&&\pyshacl &892 &0.51h&3.02s&0 &608&(40.5\%)\\
 &&&\topbraid &0 &0.81h&1.93s&0 &1500&(100.0\%)\\
\midrule\midrule
\multirow{4}{*}{\hermit}&\multirow{4}{*}{51.4h}&\multirow{4}{*}{123.3s}
 &\isaitb &1 &0.71h&1.98s&200 &1299&(86.6\%)\\
 &&&\jena &0 &0.49h&1.36s&201 &1299&(86.6\%)\\
 &&&\pyshacl &40 &0.28h&0.81s&201 &1259&(83.9\%)\\
 &&&\topbraid &1 &0.49h&1.36s&200 &1299&(86.6\%)\\
\midrule
\multirow{4}{*}{\jfact}&\multirow{4}{*}{158.2h}&\multirow{4}{*}{379.7s}
 &\isaitb &0 &0.32h&1.95s&902 &598&(39.9\%)\\
 &&&\jena &1 &0.17h&1.02s&902 &597&(39.8\%)\\
 &&&\pyshacl &1 &0.06h&0.36s&901 &598&(39.9\%)\\
 &&&\topbraid &0 &0.20h&1.23s&902 &598&(39.9\%)\\
\midrule
\multirow{4}{*}{\pellet}&\multirow{4}{*}{43.7h}&\multirow{4}{*}{105.0s}
 &\isaitb &0 &0.72h&1.97s&185 &1315&(87.7\%)\\
 &&&\jena &1 &0.50h&1.38s&186 &1313&(87.5\%)\\
 &&&\pyshacl &47 &0.21h&0.60s&186 &1267&(84.5\%)\\
 &&&\topbraid &0 &0.48h&1.31s&186 &1314&(87.6\%)\\
\midrule\midrule
\multirow{2}{*}{\pyshacl}&\multicolumn{3}{l|}{+ OWL-RL} &107 &14.8h&38.3s&0 &1393&(92.9\%)\\
 &\multicolumn{3}{l|}{+ OWL-RL + RDFS} &107 &20.8h&53.7s&0 &1393&(92.9\%)\\
\bottomrule
\end{tabular}

%% file: sections/08b-towards-full-OWL-EL.tex
\section{Discussion: towards OWL EL}
\label{sec:fullel}

Before, we considered \owlelm, a fragment of OWL EL in which existential restrictions on the right hand side of axioms are disallowed. In this section, we go over the ideas and challenges involved in converting our method into one for full OWL EL. The main issues to consider are handling the introduction of new edges to already existing nominals and the handling of the introduction of edges for which the object is not specified in the axiom. We will discuss them in this order.

\paragraph{Existential restrictions to nominals.} 
When encountering axioms of the form $C \isa \exists p.c$, this means we have to simulate jumps from nodes in the extension of the concept $C$ to the nominal $c$ whenever $p$ occurs in a shape expression. Consider for instance the constraint $s \gets \ggat p.\varphi$ with target $s(a)$. A possible way to rewrite this would be by using some kind of universal role $u$ that connects every node to every other node:
$$
s \gets \ggat p.\varphi \sqcup (\ggae p.\varphi \sqcap C \sqcap \lnot\!\ggae p.c \;\sqcap \ggae u.(c \land \varphi)).
$$
However, such a universal role is in general not included in validators, meaning that we have to simulate the described setting in some way. Here, we may use that this is not a random jump through the data graph; each one of them is initiated by ontology axioms. That is, we may include the triple structure used to define the axiom $C \isa \exists p.c$ in the rewritten shape constraint:\footnote{We omit prefixes for readability.} the structure $\mathit{type}(a,C)$,  $\mathit{subclassOf}(C, x)$, $\mathit{someValuesFrom}(x,y)$, $\mathit{oneOf}(y,z)$, $\mathit{first}(z,c)$ might be there, in which case, $u=\mathit{type}\cdot \mathit{subclassOf} \cdot \mathit{someValuesFrom} \cdot \mathit{oneOf} \cdot \mathit{first}$ will suffice. However, this gets more complicated if $a$ being of type $C$ is something that will need to be derived as well. %This is then complicated even more by axioms of the form $\rang{r} \isa A$, in which $r$ is appearing as a subject, instead of as a property.

Thus, we need to somehow move the validation process to the node $c$, but make it depend on the right conditions being met at the node $a$. 
Another way to achieve this would be to use Boolean combinations of targets: $s(a) \lor (s'(a) \land s''(c))$, where $s \gets \ggat p.\varphi$ is the original constraint, and $s'$ and $s''$ are given as follows
\begin{align*}
    &s' \gets \ggae p.\varphi \sqcap C \sqcap \lnot\!\ggae p.c & &s'' \gets \varphi.
\end{align*}

At the time of writing, there is a W3C working group developing SHACL 1.2 \cite{shacl12}, an updated  SHACL version, which among other things extends the targetting by so-called shape targets. These make it possible to target all nodes adhering to a specified shape, but they do not support boolean combinations of targets on different shapes as exemplified and needed above. That is, it remains unclear whether such an approach, although natural, will become supported in the near future.

\paragraph{Existential restrictions.}
Considering unrestricted existential restrictions in the setting of combining SHACL with reasoning has been the topic of the works by Ahmetaj et al. and Oudshoorn et al. \cite{ecai2023,dl2024,aij2026}, although in fragments without nominals.\footnote{Namely the description logics DL-Lite$_R$ (DL underlying OWL QL), \elhi\ and Horn-$\mathcal{ALCHIQ}$.} That is, rewriting techniques for SHACL and existential restrictions are known, but also known to be \exptime-complete in combined complexity (data graph size plus ontology size), for relatively inexpressive ontologies like OWL QL already \cite{ecai2023,aij2026}. The main reason for this exponential blow-up lays in having to consider every possible configuration that can appear in the data - which can only be resolved by considering which settings may actively appear in the data \cite{dl2024}, at the cost of leaving the data independence behind. That is, it is not per se impossible to perform SHACL rewritings with unrestricted existential restrictions, but it requires care to obtain efficient rewritings.

%% file: sections/09-conclusions-and-outlook.tex
% !TEX root = ../main.tex
\section{Conclusion and Outlook}\label{sec:conc}

In this work we presented the first step from the so far rather theoretical research around ontology aware shapes rewriting towards its practical application. We identified \owlelm as an OWL EL fragment which is expressive enough to cover non trivial use cases \cite{riskman} but which can still be incorporated into SHACL shapes with rather low effort. We detailed how the rewriting can be done and provided an implementation which can be reused in other work. We performed an evaluation and compared the rewriting approach to (1) the approach performing reasoning on a data graph to then validate the shapes on the resulting materialisation using different engines and (2) to a validator with built-in reasoning. We showed that the combination of rewriting and validation could solve all our test cases and was faster than the alternatives in most cases. Moreover, our rewritten shapes can be reused for new data graphs that come with the same ontology axioms and constraints further reducing validation times. Our evaluation also revealed a potential problem, namely that the rewritten shapes get rather complex. We do not expect users to directly modify these automatically generated files. However, depending on the engine, this complexity will also be visible in the validation reports. This will make reports harder to read, but allow users to better trace the cause for a violation. Future work will help to generate extracts from such complex reports optimised for the users' needs.
%which might influence the validation reports and impedes debugging. We aim to tackle this problem in future work.
%In this work, we presented the theoretical underpinnings for how to incorporate ontology axioms into SHACL constraints, making the latter ontology-aware. We bridged this theory to practice by extending known rewriting techniques and outlining an implementable procedure. A question about efficiency remains: whether the new methodology can outperform a traditional inference-and-validation setup. We conjecture that this may be the case, due to the often cumbersome nature of materialisation-based reasoners, which often times compute irrelevant conclusions and thus introduce avoidable overhead.
%which is what we plan to tackle in future work.
We furthermore want to extend our techniques to cover more expressive fragments of OWL EL by adding, for example, inverse roles. As our SHACL fragment already contains inverse roles, we expect that this generalisation could follow rather straightforwardly after a careful inspection of the described algorithms.

%% file: sections/acknowledgements.tex
% !TEX root = ../main.tex
\paragraph*{Acknowledgements.}
\imge \quad The project leading to this application has received funding from the European Union's Horizon 2020 research and innovation programme under grant agreement No 101034440. 

Furthermore, this work was supported by funding from BMFTR within projects SEMECO (grant no. 03ZU1210B),
KIMEDS (grant no. GW0552B), and MEDGE (grant no. 16ME0529).

%% file: sections/supplemental-material-statement.tex
\paragraph*{Supplemental Material Statement.}
The Turtle encoding of~\Cref{ex:running-example-declarative} (input and output of \owlawareshacl), pseudocodes of the benchmark generator, and benchmark statistics (positive-to-negative ratio, triple counts for benchmarks, materialised datagraphs, and rewritten shapes) are provided in the appendix. 
\owlawareshacl's source code, is available from GitHub at \url{https://github.com/gorczyca/owl-aware-shacl}. Additional materials, including the Java OWL API wrappers for the reasoners, the benchmark generator, result summaries in CSV format, sample benchmarks for each difficulty level (1--15), scripts for setting up and running the experiments, and scripts for reproducing~\Cref{tab:results} and~\Cref{fig:cactus-plot}, are available from GitHub at \url{https://github.com/gorczyca/owl-aware-shacl-supplementary-material}.
The full benchmark set (1{,}500 instances) and raw experimental outputs are available from Figshare under a permanent DOI: \url{https://doi.org/10.6084/m9.figshare.33169694}.

%% file: sections/use-of-gen-ai-statement.tex
\paragraph*{Use of Generative AI.}
Claude (Sonnet 4.6, Anthropic) through the Claude Code VS Code extension was used to aid in the implementation of parts of the produced code, by prompting it to provide, modify, or adjust code snippets, which were then manually reviewed and modified after careful inspection. In a similar manner, Claude aided in creating scripts for running experiments on a cluster, evaluating results, and generating summaries as seen in the paper's tables and plots, all under careful supervision and with multiple reproductions. It was additionally used for minor text editing in Section \ref{sec:implementationandevaluation}.

%% file: sections/appendix.tex
% !TEX root = ../main.tex
\section*{Appendix}
\paragraph*{Turtle Encoding.}
% \subsection*{Turtle Encoding}
\Cref{fig:running-example-input} (\Cref{fig:running-example-output}) contains the input (output) of our rewriting algorithm in Turtle format. It corresponds to the logic-based encoding from ~\Cref{ex:running-example-declarative}, helping to establish the relationship between these two notations. For example, the shape $s_2(\exists r^-.\top)$ corresponds to the triple \lstinline[style=turtle,basicstyle=\ttfamily,frame=none]{:s2 sh:targetObjectsOf :r.} (\Cref{fig:running-example-input}, right, \Cref{line:s2-target}) On the other hand, the universal targetting of bridge shapes $s_1^{\Join}(\top)$  and $s_2^{\Join}(\top)$ is obtained by addressing all named individuals, instances of all class names, subjects and objects of all role names present in the ontology as shown in~\Cref{fig:running-example-output}, Lines~\ref{line:s1-bridge} and~\ref{line:s2-bridge}. For this, and other computational reasons names of all ontology individuals, roles and classes have to be declared as in (\Cref{fig:running-example-input}, Lines~\ref{line:decl-start}--\ref{line:decl-end}).
\begin{figure}[h]
    \makebox[\textwidth][c]{%
        \begin{minipage}{1.3\textwidth}
            \begin{minipage}[t]{0.54\linewidth}
\begin{lstlisting}[style=turtle,basicstyle=\ttfamily\scriptsize]
@prefix : <http://example.org/> .
@prefix owl: <http://www.w3.org/2002/07/owl#> .
@prefix rdfs: <http://www.w3.org/2000/01/rdf-schema#> .
@prefix rdf: <http://www.w3.org/1999/02/22-rdf-syntax-ns#> .

# Classes
:A a owl:Class . (*@\label{line:decl-start}@*)
:B a owl:Class .
:C a owl:Class .
# Object properties
:r  a owl:ObjectProperty .
:r0 a owl:ObjectProperty .
:r1 a owl:ObjectProperty . (*@\label{line:decl-end}@*)
# (*@\color{Gray}$A \sqsubseteq B$@*)
:A rdfs:subClassOf :B .
# (*@\color{Gray}$\exists r.B \sqsubseteq C$@*)
[ a owl:Restriction ;
  owl:onProperty :r ;
  owl:someValuesFrom :B ] rdfs:subClassOf :C .
# (*@\color{Gray}$\dom{r} \sqsubseteq A$@*)
:r rdfs:domain :A .
# (*@\color{Gray}$\rang{r} \sqsubseteq B$@*)
:r rdfs:range :B .
# (*@\color{Gray}$r_0 \sqsubseteq r$@*)
:r0 rdfs:subPropertyOf :r .
# (*@\color{Gray}$r_1 \circ r \sqsubseteq r$@*)
:r owl:propertyChainAxiom ( :r1 :r ) .
# (*@\color{Gray}$r_0 \sqsubseteq r_1$@*)
:r0 rdfs:subPropertyOf :r1 .
\end{lstlisting}
            \end{minipage}%
            \hspace{0.02\linewidth}%
            \begin{minipage}[t]{0.44\linewidth}
\begin{lstlisting}[style=turtle,basicstyle=\ttfamily\scriptsize]
@prefix : <http://example.org/> .
@prefix sh: <http://www.w3.org/ns/shacl#> .

# (*@\color{Gray}$s_1 \gets \exists r.B,\quad s_1(A)$@*)
:s1 a sh:NodeShape ;
    sh:targetClass :A ;
    sh:property [
        sh:path :r ;
        sh:class :B ;
        sh:minCount 1
    ] .

# (*@\color{Gray}$s_2 \gets \exists r^-.A,\quad s_2(\exists r^-.\!\top)$@*)
:s2 a sh:NodeShape ;
    sh:targetObjectsOf :r ; (*@\label{line:s2-target}@*)
    sh:property [
        sh:path [ sh:inversePath :r ] ;
        sh:class :A ;
        sh:minCount 1
    ] .
\end{lstlisting}
            \end{minipage}
        \end{minipage}}
    \caption{Input to the implementation, Turtle encoding of the example from \Cref{ex:running-example-declarative}: TBox $\TBox$ (left) and shapes $\C$ 
    % $\C,\G$ 
    (right).
    }
    \label{fig:running-example-input}
\end{figure}
\begin{figure}[p]
    \makebox[\textwidth][c]{%
        \begin{minipage}{1.3\textwidth}
\begin{lstlisting}[style=turtle,basicstyle=\ttfamily\scriptsize]
@prefix ex: <http://example.org/> .
@prefix rdf: <http://www.w3.org/1999/02/22-rdf-syntax-ns#> .
@prefix s: <https://example.org/shapes#> .
@prefix sh: <http://www.w3.org/ns/shacl#> .
@prefix xsd: <http://www.w3.org/2001/XMLSchema#> .

# (*@\color{Gray}$s_1 \gets \exists\bigl((r_0 \mid r_1)^* \cdot (r_0 \mid r)\bigr).s_B$@*)
ex:s1 a sh:NodeShape ;
    sh:property [ sh:minCount 1 ;
            sh:node s:S_CLASS_B ;
            sh:path ( [ sh:zeroOrMorePath [ sh:alternativePath ( ex:r0 ex:r1 ) ] ] [ sh:alternativePath ( ex:r ex:r0 ) ] ) ] ;
    sh:targetClass ex:A .
# (*@\color{Gray}$s_2 \gets \exists\bigl((r_0^- \mid r^-) \cdot (r_0^- \mid r_1^-)^*\bigr).s_A$@*)
ex:s2 a sh:NodeShape ;
    sh:property [ sh:minCount 1 ;
            sh:node s:S_CLASS_A ;
            sh:path ( [ sh:alternativePath ( [ sh:inversePath ex:r ] [ sh:inversePath ex:r0 ] ) ] [ sh:zeroOrMorePath [ sh:alternativePath ( [ sh:inversePath ex:r0 ] [ sh:inversePath ex:r1 ] ) ] ] ) ] ;
    sh:targetObjectsOf ex:r .
# (*@\color{Gray}$s_A \gets A \lor s_{\exists r.\top}$@*)
s:S_CLASS_A a sh:NodeShape ;
    sh:node [ sh:or ( s:S_DomainOf_r [ sh:class ex:A ] ) ] .
# (*@\color{Gray}$s_B \gets B \lor s_A \lor s_{\exists r^-.\top}$@*)
s:S_CLASS_B a sh:NodeShape ;
    sh:node [ sh:or ( s:S_CLASS_A s:S_RangeOf_r [ sh:class ex:B ] ) ] .
# (*@\color{Gray}$s_C \gets C \lor s_{\exists r.B}$@*)
s:S_CLASS_C a sh:NodeShape ;
    sh:node [ sh:or ( s:S_EXISTS_r_CLASS_B [ sh:class ex:C ] ) ] .
# (*@\color{Gray}$s_1^{\Join} \gets \neg s_A \lor s_1,\quad s_1^{\Join}(\top)$@*)
s:s1_Bridge a sh:NodeShape ; (*@\label{line:s1-bridge}@*)
    sh:node [ sh:or ( [ sh:not [ sh:or ( s:S_CLASS_A ) ] ] ex:s1 ) ] ;
    sh:targetClass ex:A, ex:B, ex:C ;
    sh:targetObjectsOf ex:r, ex:r0, ex:r1 ;
    sh:targetSubjectsOf ex:r, ex:r0, ex:r1 .
# (*@\color{Gray}$s_2^{\Join} \gets \neg s_{\exists r^-.\top} \lor s_2,\quad s_2^{\Join}(\top)$@*)
s:s2_Bridge a sh:NodeShape ; (*@\label{line:s2-bridge}@*)
    sh:node [ sh:or ( [ sh:not [ sh:or ( s:s2_TargetObjectsOf_r ) ] ] ex:s2 ) ] ;
    sh:targetClass ex:A, ex:B, ex:C ;
    sh:targetObjectsOf ex:r, ex:r0, ex:r1 ;
    sh:targetSubjectsOf ex:r, ex:r0, ex:r1 .
# (*@\color{Gray}$s_{\exists r.\top} \gets \exists\bigl((r_0 \mid r_1)^* \cdot (r_0 \mid r)\bigr).\top$@*)
s:S_DomainOf_r a sh:NodeShape ;
    sh:property [ sh:minCount 1 ;
            sh:path ( [ sh:zeroOrMorePath [ sh:alternativePath ( ex:r0 ex:r1 ) ] ] [ sh:alternativePath ( ex:r ex:r0 ) ] ) ] .
# (*@\color{Gray}$s_{\exists r^-.\top} \gets \exists\bigl((r_0^- \mid r^-) \cdot (r_0^- \mid r_1^-)^*\bigr).\top$@*)
s:S_RangeOf_r a sh:NodeShape ;
    sh:property [ sh:minCount 1 ;
            sh:path ( [ sh:alternativePath ( [ sh:inversePath ex:r ] [ sh:inversePath ex:r0 ] ) ] [ sh:zeroOrMorePath [ sh:alternativePath ( [ sh:inversePath ex:r0 ] [ sh:inversePath ex:r1 ] ) ] ] ) ] .
# (*@\color{Gray}$s_{\exists r.B} \gets \exists\bigl((r_0 \mid r_1)^* \cdot (r_0 \mid r)\bigr).s_B$@*)
s:S_EXISTS_r_CLASS_B a sh:NodeShape ;
    sh:property [ sh:path ( [ sh:zeroOrMorePath [ sh:alternativePath ( ex:r0 ex:r1 ) ] ] [ sh:alternativePath ( ex:r ex:r0 ) ] ) ;
            sh:qualifiedMinCount 1 ;
            sh:qualifiedValueShape s:S_CLASS_B ] .
# (*@\color{Gray} helper shape for target of $s_2$, evaluates to the same as $s_{\exists r^-.\top}$, because objects of $r$ is the same as range of $r$ @*)
s:s2_TargetObjectsOf_r a sh:NodeShape ;
    sh:property [ sh:minCount 1 ;
            sh:path ( [ sh:alternativePath ( [ sh:inversePath ex:r ] [ sh:inversePath ex:r0 ] ) ] [ sh:zeroOrMorePath [ sh:alternativePath ( [ sh:inversePath ex:r0 ] [ sh:inversePath ex:r1 ] ) ] ] ) ] .
\end{lstlisting}
        \end{minipage}}
    \caption{Turtle output of the implementation for the example from \Cref{ex:running-example-declarative}, given the input from~\Cref{fig:running-example-input}.}
    \label{fig:running-example-output}
\end{figure}
\paragraph*{Benchmarks Generation.}
Throughout this subsection, parameter names are typeset in \textcolor{teal}{\textit{teal italics}} (e.g., \param{numClasses}) to make them easily distinguishable across the table and algorithms.
\Cref{tab:benchmark} lists all parameters of the benchmark generator together with their descriptions, and provides the concrete instantiations of each parameter across the 15 difficulty levels used in the evaluation.
The benchmarks are generated by three algorithms.
\textsc{GenerateTBox} (\Cref{alg:tbox}) produces an OWL EL$^-$ TBox $\mathcal{T}$ along with the sets of class names $\NC$, role names $\NR$, and two set of individuals $\NI$, $\NA$.
\textsc{GenerateDataGraph} (\Cref{alg:abox}) populates a data graph $\mathcal{A}$ with random class and role assertions over the data graph individuals $\NA$, class names $\NC$, and role names $\NR$.
\textsc{GenerateSHACL} (\Cref{alg:shacl}) produces a set of SHACL shapes $\C$ with targets $\mathcal{G}$, using randomly generated property paths via the auxiliary \textsc{GeneratePath} function; as \param{pathContinueProbability} increases with the difficulty level, higher levels yield more deeply nested and structurally complex paths.
Each difficulty level is instantiated with 100 benchmark instances, yielding 1{,}500 instances in total.

\input{sections/appendix/benchmark-tables}
\input{sections/appendix/pseudocode-preamble}
\input{sections/appendix/alg-tbox}
\input{sections/appendix/alg-abox}
\input{sections/appendix/alg-shacl}
\paragraph*{Benchmarks Statistics. }
\Cref{tab:stats} reports statistics for each of the 15 difficulty levels across the 1{,}500 generated instances.
Since both the ontology and the shapes are randomly generated, it is unlikely that the data will conform to the shapes, and indeed most instances are non-conforming (negative). Level~1 is an exception with an exactly even 50/50 split; from level~2 onward virtually all instances are negative.
The table also reports mean RDF triple counts per instance for the input TBox (\texttt{owl.ttl}), the input SHACL shapes (\texttt{shacl.ttl}), and the rewritten output (\texttt{rewriting.ttl}), to help get an idea of how large the output rewriting becomes with respect to the input ontology and shapes.
As expected, triple counts grow with the difficulty level.
\input{sections/evaluation/benchmark-evaluation-table}

\paragraph*{Extra Results.}
\Cref{tab:results-min-med-max} complements \Cref{tab:results} by reporting, for the same full instance set, the minimum, median, and maximum reasoning/rewriting and validation time per approach and validator, computed over successfully completed instances only.
\Cref{tab:results-intersection-no-jfact} 
shows evaluation results restricted to instances solved by every approach, but excluding \jfact for the baseline method and \pyshacl for rewriting, since both are unable to solve too many instances. \Cref{tab:results-intersection-no-jfact-min-med-max} reports the corresponding minimum, median, and maximum timing statistics for this same restricted instance set.

\begin{table}[h]
    \centering
    \input{sections/evaluation/results_table_min_med_max}
    \vspace{10pt}
    \caption{Minimum, median, and maximum reasoning/rewriting time (Min./Med./Max., left) and validation time (Min./Med./Max., right) per approach and validator, computed over successfully completed instances only (i.e., excluding reasoning/rewriting timeouts and validator errors). Note that the minimum validation times are nearly identical between validators. This is likely due to the overhead of the validator itself per call, i.e., a fixed cost of launching the validator as a fresh process. The minima do in fact differ slightly between validators (by a few milliseconds), but this difference is lost due to rounding.}
    \label{tab:results-min-med-max}
\end{table}

\begin{table}[h]
    \centering
    \input{sections/evaluation/results_table_intersection_no_jfact}
    \vspace{10pt}
    \caption{Results per approach and validator, restricted to instances solved by every approach excluding \jfact for the baseline method and \pyshacl for rewriting.}
    \label{tab:results-intersection-no-jfact}
\end{table}

\begin{table}[h]
    \centering
    \input{sections/evaluation/results_table_intersection_no_jfact_min_med_max}
    \vspace{10pt}
    \caption{Minimum, median, and maximum reasoning/rewriting and validation time per approach and validator, restricted to the same instance set as \Cref{tab:results-intersection-no-jfact} (solved by every approach, excluding \jfact for the baseline method and \pyshacl for rewriting), computed over successfully completed instances only.}
    \label{tab:results-intersection-no-jfact-min-med-max}
\end{table}

%% file: sections/appendix/benchmark-tables.tex
\begin{table}[h]
    \centering
    \begin{tabularx}{\linewidth}{lX}
        \toprule
        \textbf{Parameter} & \textbf{Description} \\
        \midrule
        \param{numClasses} (nC)              & Number of named classes (class names) \\
        \param{numRoles} (nR)                & Number of object properties (role names) \\
        \param{numIndividuals} (nI)          & Total individuals; first 20\% are nominals ($\NI$), rest are data graph individuals ($\NA$) \\
        \param{numGCIs} (nGCI)               & Number of general concept inclusions \\
        \param{numDomains} (nD)              & Number of domain axioms $\dom{r} \sqsubseteq C$ \\
        \param{numRanges} (nRn)              & Number of range axioms $\ran{r} \sqsubseteq C$ \\
        \param{numRoleChains} (nCh)          & Number of role chain axioms $r_i \circ r_j \sqsubseteq r_j$ \\
        \param{numRoleInclusions} (nInc)     & Number of role inclusion axioms $r_i \sqsubseteq r_j$ \\
        \param{avgClassAssertions} (aCA)     & Average class assertions per individual in the data graph \\
        \param{avgRoleAssertions} (aRA)      & Average role assertions per individual in the data graph \\
        \param{numShapes} (nS)               & Number of SHACL shapes to generate \\
        \param{pathContinueProbability} (pCP)& Probability of extending a property path recursively \\
        \bottomrule
    \end{tabularx}

    \bigskip

    \setlength{\tabcolsep}{4pt}
    \small
    \begin{tabular}{r|rrrrrrrrrrrr}
        \toprule
        Level & nC & nR & nI & nGCI & nD & nRn & nCh & nInc & aCA & aRA & nS & pCP \\
        \midrule
        1  &      4 &     2 &     10 &      4 &   2 &   2 &   1 &    1 & 2.5 &  3.0 &  2 & 0.30 \\
        2  &     11 &     5 &     20 &     10 &   2 &   2 &   1 &    1 & 2.5 &  3.1 &  2 & 0.31 \\
        3  &     17 &     8 &     35 &     22 &   2 &   2 &   1 &    2 & 2.5 &  3.2 &  2 & 0.32 \\
        4  &     24 &    12 &     50 &     40 &   2 &   2 &   2 &    4 & 2.5 &  3.5 &  2 & 0.35 \\
        5  &     50 &    25 &    100 &     80 &   5 &   5 &   4 &    8 & 3.0 &  4.0 &  3 & 0.40 \\
        6  &     80 &    40 &    150 &    130 &   8 &   8 &   6 &   12 & 3.0 &  4.5 &  4 & 0.45 \\
        7  &    120 &    60 &    250 &    200 &  12 &  12 &   8 &   18 & 3.5 &  5.0 &  4 & 0.50 \\
        8  &    180 &    90 &    400 &    300 &  18 &  18 &  12 &   25 & 3.5 &  5.5 &  5 & 0.55 \\
        9  &    250 &   125 &    600 &    450 &  25 &  25 &  16 &   35 & 4.0 &  6.0 &  5 & 0.60 \\
        10 &    350 &   175 &    900 &    650 &  35 &  35 &  22 &   50 & 4.0 &  6.5 &  6 & 0.65 \\
        11 &    500 &   250 &  1{,}300 &    950 &  50 &  50 &  30 &   70 & 4.5 &  7.0 &  7 & 0.65 \\
        12 &    700 &   350 &  2{,}000 &  1{,}400 &  70 &  70 &  40 &  100 & 5.0 &  8.0 &  8 & 0.70 \\
        13 &  1{,}000 &   500 &  3{,}000 &  2{,}000 & 100 & 100 &  55 &  140 & 5.0 &  9.0 &  9 & 0.70 \\
        14 &  1{,}500 &   750 &  4{,}500 &  3{,}000 & 150 & 150 &  75 &  200 & 5.5 & 10.0 & 10 & 0.75 \\
        15 &  2{,}200 & 1{,}100 &  6{,}500 &  4{,}500 & 220 & 220 & 100 &  280 & 6.0 & 11.0 & 11 & 0.75 \\
        \bottomrule
    \end{tabular}
    \vspace{10pt}
    \caption{Benchmark generator parameters (top) and their instantiations across difficulty levels 1--15 (bottom).}
    \label{tab:benchmark}
\end{table}

%% file: sections/appendix/pseudocode-preamble.tex
\LinesNumbered
\SetKwProg{Fn}{Function}{:}{}
\SetKwFunction{GenerateTBox}{GenerateTBox}
\SetKwFunction{GenerateABox}{GenerateABox}
\SetKwFunction{GenerateSHACL}{GenerateSHACL}
\SetKwFunction{GeneratePath}{GeneratePath}
\SetKwFunction{randomRole}{randomRole}
\SetKwFunction{random}{random}
\SetKwComment{tcc}{// }{}
\DontPrintSemicolon

%% file: sections/appendix/alg-tbox.tex
\begin{algorithm}[h]
\caption{\textsc{GenerateTBox}}\label{alg:tbox}
\KwIn{\param{numClasses}, \param{numRoles}, \param{numGCIs}, \param{numDomains}, \param{numRanges}, \param{numRoleChains}, \param{numRoleInclusions}, \param{numIndividuals}}
\KwOut{OWL EL$^-$ TBox $\mathcal{T}$; sets $\NC$, $\NR$, $\NI$, $\NA$}
$\NC \leftarrow \{C_1,\ldots,C_{\param{numClasses}}\}$ with index $\mathit{idx}(C_i) = i$\;
$\NR \leftarrow \{r_1,\ldots,r_{\param{numRoles}}\}$ with index $\mathit{idx}(r_i) = i$\;
$\NI' \leftarrow \{a_1,\ldots,a_{\param{numIndividuals}}\}$\;
$\NI \leftarrow$ first $\lfloor 0.2\cdot|\NI'|\rfloor$ elements of $\NI'$\;
$\NA \leftarrow \NI' \setminus \NI$\;
$\mathcal{T} \leftarrow \emptyset$\;
\BlankLine
\tcc{GCI generation}
$n \leftarrow 0$\;
\While{$n < \param{numGCIs}$}{
  $L \leftarrow$ random LHS from $\{A,\ A\sqcap B,\ \exists r.A,\ A\sqcap\exists r.B,\ \exists r.\{a\},\ A\sqcap\exists r.\{a\}\}$ where $A,B\in\NC$, $r\in\NR$, $a\in\NI$\;
  $B \leftarrow \random(\NC)$\;
  \If{$\mathit{idx}(C) < \mathit{idx}(B)$ for every class $C$ appearing in $L$}{
    add $L \sqsubseteq B$ to $\mathcal{T}$\;
    $n \leftarrow n + 1$\;
  }
}
\BlankLine
\tcc{Domain/range axioms}
\ForEach{$r$ in \param{numDomains} randomly chosen elements of $\NR$}{
  add $\mathit{dom}(r) \sqsubseteq \random(\mathcal{C})$ to $\mathcal{T}$\;
}
\ForEach{$r$ in \param{numRanges} randomly chosen elements of $\NR$}{
  add $\mathit{rang}(r) \sqsubseteq \random(\mathcal{C})$ to $\mathcal{T}$\;
}
\BlankLine
\tcc{Role chain axioms}
$n \leftarrow 0$\;
\While{$n < \param{numRoleChains}$}{
  pick $r_i, r_j \in \NR$ randomly\;
  \If{$i < j$}{
    add $r_i \circ r_j \sqsubseteq r_j$ to $\mathcal{T}$\;
    $n \leftarrow n + 1$\;
  }
}
\BlankLine
\tcc{Role inclusions}
$n \leftarrow 0$\;
\While{$n < \param{numRoleInclusions}$}{
  pick $r_i, r_j \in \NR$ randomly\;
  \If{$i < j$}{
    add $r_i \sqsubseteq r_j$ to $\mathcal{T}$\;
    $n \leftarrow n + 1$\;
  }
}
\BlankLine
\Return $\mathcal{T}$, $\NC$, $\NR$, $\NI$, $\NA$\;
\end{algorithm}

%% file: sections/appendix/alg-abox.tex
\begin{algorithm}[h]
\caption{\textsc{GenerateDataGraph}}\label{alg:abox}
\KwIn{\param{numIndividuals}, \param{avgClassAssertions}, \param{avgRoleAssertions}}
\KwOut{Data Graph $\mathcal{A}$}
$\mathcal{A} \leftarrow \emptyset$\;
\BlankLine
\tcc{Class assertions}
$n \leftarrow 0$\;
\While{$n < \lfloor \param{numIndividuals}\cdot\param{avgClassAssertions} \rfloor$}{
  $C \leftarrow \random(\NC)$\;
  $i \leftarrow \random(\NA)$\;
  add $C(i)$ to $\mathcal{A}$\;
  $n \leftarrow n + 1$\;
}
\BlankLine
\tcc{Role assertions}
$n \leftarrow 0$\;
\While{$n < \lfloor \param{numIndividuals}\cdot\param{avgRoleAssertions} \rfloor$}{
  $r \leftarrow \random(\NR$\;
  $i \leftarrow \random(\NA)$\;
  $j \leftarrow \random(\NA)$\;
  add $r(i, j)$ to $\mathcal{A}$\;
  $n \leftarrow n + 1$\;
}
\BlankLine
\Return $\mathcal{A}$\;
\end{algorithm}

%% file: sections/appendix/alg-shacl.tex
\begin{algorithm}[h]
\caption{\textsc{GenerateSHACL}}\label{alg:shacl}
\KwIn{\param{numShapes}, \param{pathContinueProbability}, $\NC$, $\NR$, $\NA$}
\KwOut{SHACL shape set $\C$, target set $\mathcal{G}$}
$\C \leftarrow \emptyset$\;
$\mathcal{G} \leftarrow \emptyset$\;
$n \leftarrow 0$\;
\BlankLine
\While{$n < \param{numShapes}$}{
  $\mathit{kind} \leftarrow \random(\{\texttt{class},\ \texttt{node},\ \texttt{subjects},\ \texttt{objects}\})$\;
  $\mathit{path} \leftarrow \GeneratePath{}$\;
  \BlankLine
  \If{$\mathit{kind} = \texttt{class}$}{
    $C \leftarrow \random(\NC)$\;
    $C' \leftarrow \random(\NC)$\;
    $s \gets \exists\,\mathit{path}.\,C'$\;
    add $s$ to $\C$\; add $s(C)$ to $\mathcal{G}$\;
  }
  \If{$\mathit{kind} = \texttt{node}$}{
    $i \leftarrow \random(\NA)$\;
    $s \gets \exists\,\mathit{path}.\top$\;
    add $s$ to $\C$\; add $s(i)$ to $\mathcal{G}$\;
  }
  \If{$\mathit{kind} = \texttt{subjects}$}{
    $r \leftarrow \random(\NR)$\;
    $s \gets \exists\,\mathit{path}.\top$\;
    add $s$ to $\C$\; add $s(\exists r.\top)$ to $\mathcal{G}$\;
  }
  \If{$\mathit{kind} = \texttt{objects}$}{
    $r \leftarrow \random(\NR)$\;
    $C \leftarrow \random(\NC)$\;
    $s \gets C$\;
    add $s$ to $\C$\; add $s(\exists r^-.\top)$ to $\mathcal{G}$\;
  }
  $n \leftarrow n + 1$\;
}
\BlankLine
\Return $\C$, $\mathcal{G}$\;
\BlankLine
\Fn{\GeneratePath{}}{
  $\mathit{kind} \leftarrow \random(\{\texttt{simple},\ \texttt{sequence},\ \texttt{alternative},\ \texttt{inverse}\})$\;
  pick $p \in [0,1]$ uniformly\;
  \If{$p > \param{pathContinueProbability}$ \textbf{or} $\mathit{kind} = \texttt{simple}$}{
    $r \leftarrow \random(\NR)$\;
    \Return $r$\;
  }
  \lIf{$\mathit{kind} = \texttt{sequence}$}{\Return $\GeneratePath{}\cdot\GeneratePath{}$}
  \lIf{$\mathit{kind} = \texttt{alternative}$}{\Return $\GeneratePath{}\mid\GeneratePath{}$}
  \lIf{$\mathit{kind} = \texttt{inverse}$}{\Return $\GeneratePath{}^-$}
}
\end{algorithm}

%% file: sections/evaluation/benchmark-evaluation-table.tex
\begin{table*}[t]
    \centering
    \setlength{\tabcolsep}{4pt}
    \begin{tabular}{l|r r | r r r | r r r r}
        \toprule
        \multirow{2}{*}{Level} & \multicolumn{2}{c|}{\#instances} & \multicolumn{7}{c}{mean \#triples per instance} \\
        \cmidrule(lr){2-3}\cmidrule(lr){4-10}
        & Pos. & Neg. & TBox & Shapes & Data Graph & Rewriting & \hermit & \jfact & \pellet \\
        \midrule
        1  & 50 & 50  & 44         & 12  & 55          & 196        & 135        & 135        & 134        \\
        2  & 4  & 96  & 97         & 12  & 112         & 302        & 242        & 242        & 242        \\
        3  & 0  & 100 & 193        & 12  & 199         & 519        & 420        & 419        & 421        \\
        4  & 2  & 98  & 333        & 12  & 300         & 849        & 669        & 664        & 678        \\
        5  & 0  & 100 & 666        & 19  & 700         & 1\,789     & 1\,498     & 1\,490     & 1\,505     \\
        6  & 0  & 100 & 1\,083     & 29  & 1\,125      & 3\,022     & 2\,380     & 2\,367     & 2\,383     \\
        7  & 0  & 100 & 1\,654     & 30  & 2\,125      & 4\,567     & 4\,370     & 4\,514     & 4\,370     \\
        8  & 0  & 100 & 2\,498     & 45  & 3\,600      & 7\,082     & 7\,267     & --         & 7\,271     \\
        9  & 0  & 100 & 3\,681     & 45  & 6\,000      & 10\,230    & 12\,049    & --         & 12\,051    \\
        10 & 0  & 100 & 5\,299     & 70  & 9\,450      & 15\,396    & 19\,083    & --         & 19\,084    \\
        11 & 0  & 100 & 7\,699     & 86  & 14\,950     & 23\,130    & 29\,894    & --         & 29\,896    \\
        12 & 0  & 100 & 11\,315    & 106 & 26\,000     & 35\,117    & 51\,491    & --         & 51\,523    \\
        13 & 0  & 100 & 16\,146    & 128 & 42\,000     & 52\,280    & 83\,256    & --         & 83\,252    \\
        14 & 0  & 100 & 24\,168    & 372 & 69\,750     & 80\,679    & --         & --         & 135\,196   \\
        15 & 0  & 100 & 36\,054    & 340 & 110\,500    & 122\,659   & --         & --         & --         \\
        \midrule
        Total & 56 & 1\,444 & 7\,395 & 88 & 19\,124 & 23\,854 & --         & --         & --         \\
        \bottomrule
    \end{tabular}
    \vspace{10pt}
    \caption{Per-level instance statistics and mean RDF triple counts. Positive/Negative: conforming/non-conforming instances. OWL: input TBox; SHACL: input shapes; Rewriting: output of the OWL-aware rewriting; \hermit/\jfact/\pellet: materialized data graph after reasoning.}
    \label{tab:stats}
\end{table*}

%% file: sections/evaluation/results_table_min_med_max.tex
\begin{tabular}{l@{\hspace{10pt}}r@{\hspace{10pt}}r@{\hspace{10pt}}r@{\hspace{14pt}}|@{\hspace{14pt}}l@{\hspace{10pt}}r@{\hspace{10pt}}r@{\hspace{10pt}}r}
\toprule
Approach&Min.&Med.&Max.&Validator&Min.&Med.&Max.\\
\midrule
\multirow{4}{*}{Rewrit.}&\multirow{4}{*}{0.18s}&\multirow{4}{*}{0.88s}&\multirow{4}{*}{15.6s}&\isaitb &1.75s&2.13s&8.52s\\
 &&&&\jena &0.93s&1.26s&6.92s\\
 &&&&\pyshacl &0.24s&0.48s&145.0s\\
 &&&&\topbraid &1.14s&1.40s&8.35s\\
\midrule\midrule
\multirow{4}{*}{\hermit}&\multirow{4}{*}{0.47s}&\multirow{4}{*}{2.29s}&\multirow{4}{*}{511.6s}&\isaitb &1.78s&1.92s&7.84s\\
 &&&&\jena &0.92s&1.04s&317.8s\\
 &&&&\pyshacl &0.29s&0.39s&270.3s\\
 &&&&\topbraid &1.14s&1.27s&41.4s\\
\midrule
\multirow{4}{*}{\jfact}&\multirow{4}{*}{0.52s}&\multirow{4}{*}{1.40s}&\multirow{4}{*}{573.1s}&\isaitb &1.78s&1.83s&42.0s\\
 &&&&\jena &0.92s&0.99s&4.24s\\
 &&&&\pyshacl &0.29s&0.32s&0.82s\\
 &&&&\topbraid &1.14s&1.19s&4.61s\\
\midrule
\multirow{4}{*}{\pellet}&\multirow{4}{*}{0.49s}&\multirow{4}{*}{1.17s}&\multirow{4}{*}{599.4s}&\isaitb &1.78s&1.89s&6.21s\\
 &&&&\jena &0.92s&1.05s&306.0s\\
 &&&&\pyshacl &0.29s&0.38s&3.01s\\
 &&&&\topbraid &1.14s&1.23s&7.26s\\
\midrule\midrule
\multirow{2}{*}{\pyshacl}&\multicolumn{4}{l}{+ OWL-RL} &0.42s&6.51s&360.7s\\
 &\multicolumn{4}{l}{+ OWL-RL + RDFS} &0.47s&10.9s&480.3s\\
\bottomrule
\end{tabular}

%% file: sections/evaluation/results_table_intersection_no_jfact.tex
\begin{tabular}{l@{\hspace{4pt}}r@{\hspace{4pt}}r@{\hspace{6pt}}|@{\hspace{6pt}}l@{\hspace{7pt}}r@{\hspace{7pt}}r@{\hspace{4pt}}r@{\hspace{4pt}}|@{\hspace{6pt}}r@{\hspace{7pt}}r@{\hspace{3pt}}r@{}}
\toprule
Approach&T.R.&Av.R.&Val.&V.E.&T.V.&Av.V.&T/O&\multicolumn{2}{c}{Succ. (\%)}\\
\midrule
\multirow{3}{*}{Rewrit.}&\multirow{3}{*}{0.44h}&\multirow{3}{*}{1.26s}
 &\isaitb &0 &0.78h&2.25s&0 &1254&(100.0\%)\\
 &&&\jena &0 &0.47h&1.36s&0 &1254&(100.0\%)\\
 &&&\topbraid &0 &0.54h&1.54s&0 &1254&(100.0\%)\\
\midrule\midrule
\multirow{4}{*}{\hermit}&\multirow{4}{*}{15.4h}&\multirow{4}{*}{44.3s}
 &\isaitb &0 &0.69h&1.97s&0 &1254&(100.0\%)\\
 &&&\jena &0 &0.47h&1.36s&0 &1254&(100.0\%)\\
 &&&\pyshacl &0 &0.28h&0.81s&0 &1254&(100.0\%)\\
 &&&\topbraid &0 &0.47h&1.36s&0 &1254&(100.0\%)\\
\midrule
\multirow{4}{*}{\pellet}&\multirow{4}{*}{9.08h}&\multirow{4}{*}{26.1s}
 &\isaitb &0 &0.68h&1.95s&0 &1254&(100.0\%)\\
 &&&\jena &0 &0.47h&1.36s&0 &1254&(100.0\%)\\
 &&&\pyshacl &0 &0.20h&0.59s&0 &1254&(100.0\%)\\
 &&&\topbraid &0 &0.45h&1.28s&0 &1254&(100.0\%)\\
\midrule\midrule
\multirow{2}{*}{\pyshacl}&\multicolumn{3}{l}{+ OWL-RL} &0 &5.80h&16.6s&0 &1254&(100.0\%)\\
 &\multicolumn{3}{l}{+ OWL-RL + RDFS} &0 &8.77h&25.2s&0 &1254&(100.0\%)\\
\bottomrule
\end{tabular}

%% file: sections/evaluation/results_table_intersection_no_jfact_min_med_max.tex
\begin{tabular}{l@{\hspace{10pt}}r@{\hspace{10pt}}r@{\hspace{10pt}}r@{\hspace{14pt}}|@{\hspace{14pt}}l@{\hspace{10pt}}r@{\hspace{10pt}}r@{\hspace{10pt}}r}
\toprule
Approach&Min.&Med.&Max.&Validator&Min.&Med.&Max.\\
\midrule
\multirow{3}{*}{Rewrit.}&\multirow{3}{*}{0.18s}&\multirow{3}{*}{0.62s}&\multirow{3}{*}{6.80s}&\isaitb &1.75s&2.02s&4.92s\\
 &&&&\jena &0.93s&1.17s&3.61s\\
 &&&&\topbraid &1.14s&1.32s&4.38s\\
\midrule\midrule
\multirow{4}{*}{\hermit}&\multirow{4}{*}{0.47s}&\multirow{4}{*}{2.22s}&\multirow{4}{*}{511.6s}&\isaitb &1.78s&1.92s&7.84s\\
 &&&&\jena &0.92s&1.03s&317.8s\\
 &&&&\pyshacl &0.29s&0.39s&270.3s\\
 &&&&\topbraid &1.14s&1.26s&41.4s\\
\midrule
\multirow{4}{*}{\pellet}&\multirow{4}{*}{0.49s}&\multirow{4}{*}{1.03s}&\multirow{4}{*}{432.7s}&\isaitb &1.78s&1.88s&6.21s\\
 &&&&\jena &0.92s&1.04s&306.0s\\
 &&&&\pyshacl &0.29s&0.37s&2.47s\\
 &&&&\topbraid &1.14s&1.22s&4.79s\\
\midrule\midrule
\multirow{2}{*}{\pyshacl}&\multicolumn{4}{l}{+ OWL-RL} &0.42s&4.59s&112.5s\\
 &\multicolumn{4}{l}{+ OWL-RL + RDFS }&0.47s&7.47s&165.0s\\
\bottomrule
\end{tabular}